\documentclass[3p,twocolumn]{elsarticle}
\usepackage{pifont}
\usepackage{natbib}
\usepackage{geometry}
\usepackage[fleqn]{amsmath}
\usepackage{graphicx}
\usepackage{txfonts}
\usepackage{hyperref}
\biboptions{sort&compress}
\pdfoutput=1

\usepackage{amssymb,amsmath}
\usepackage{multicol}
\usepackage{multirow}
\usepackage{framed}
\usepackage[usenames,dvipsnames,svgnames,table]{xcolor}
\usepackage{cancel} 
\usepackage{soul}
\usepackage{caption}
\usepackage{subcaption}
\usepackage{latexsym} 
\usepackage{mathrsfs}
\usepackage{amsxtra}

\usepackage{color}
\newcommand{\black}[1]{\textcolor{black}{#1}}
\newcommand{\red}[1]{\textcolor{red}{#1}}

\newcommand{\pd}{\partial}

\newcommand{\e}{\varepsilon}
\renewcommand{\d}{\delta}
\newcommand{\f}{\varphi}
\renewcommand{\l}{\lambda}
\newcommand{\w}{\omega}

\renewcommand{\k}{\kappa}

\newcommand{\EE}{\mathcal{E}}

\newcommand{\Fig}[1]{Figure \ref{#1}}
\newcommand{\Tab}[1]{Table \ref{#1}}

\newcommand{\MLt}[3]{E_{#1,#2}\left(#3 \right)}  
\newcommand{\MLo}[2]{E_{#1}\left(#2 \right)}       

\newcommand{\DD}[1]{\mathcal{D}^{#1}_{t}} 

\newcommand{\intO}{\int_{\Omega}}
\newcommand{\dO}[1]{\left[ #1 \right]_{\partial \Omega}}
\newcommand{\sss}[1]{\scriptscriptstyle{#1}}

\newcommand{\dd}{\mathrm{d}}

\newcommand{\bF}{\bar{\f}}
\newcommand{\bA}{\bar{\alpha}}

\newcommand{\bC}{\bar{C}}

\newcommand{\betaL}{\lambda}
\newcommand{\betaN}{\nu}

\newcommand{\psidz}{\psi_{\sss{DZ}}}
\newcommand{\Psidz}{\Psi_{\sss{DZ}}}

\newcommand{\CL}{C_{\sss{L}}}
\newcommand{\CN}{C_{\sss{N}}}
\newcommand{\bCL}{C_{\sss{L}}^*}
\newcommand{\bCN}{C_{\sss{N}}^*}

\newcommand\bx{\textbf{\emph{x}}}
\newcommand\be{\textbf{\emph{e}}}
\newcommand\bn{\textbf{\emph{n}}}
\newcommand\sx{\textbf{\textsf{x}}}
\newcommand\scrE{\mathscr{E}}
\newcommand\scrI{\mathscr{I}}
\newcommand\scrW{\mathscr{W}}
\newcommand\ff{\textbf{\emph{f}}}
\renewcommand\gg{\textbf{\emph{g}}}
\newcommand\FF{\textbf{F}}

\newcommand\grad{\text{grad}}

\newcommand\Body{\mathscr{B}}
\newcommand{\wh}{\widehat}

\journal{JMBBM}
\begin{document}
\begin{frontmatter}
\title{Fractional Hereditariness of Lipid Membranes:\\ Instabilities and Linearized Evolution}

\author[address1,address2,address3]{L. Deseri\corref{cor1}}
\ead{deseri@andrew.cmu.edu}
\author[address3]{P. Pollaci}
\ead{pietro.pollaci@unitn.it}
\author[address4,address5]{M. Zingales}
\ead{massimiliano.zingales@unipa.it}
\author[address6]{K. Dayal}
\ead{kaushik@cmu.edu}

\cortext[cor1]{Corresponding author}

\address[address1]{Civil and Environmental Engineering-CEE\\ Carnegie Mellon University, Pittsburgh PA 15213-3890, USA}

\address[address2]{TMHRI-Department of Nanomedicine, The Methodist Hospital Research Institute \\ MS B-490 Houston, TX 77030 USA}

\address[address3]{DICAM - Civil, Environmental and Mechanical Engineering \\ University of Trento, via Mesiano 77, 38123 Trento, Italy}

\address[address4]{DICAAM - Civil, Environmental, Aerospace Engineering and Material Science  \\ University of Palermo, Viale delle Science, Edificio 8, 90100 Palermo, Italy}

\address[address5]{(BM)$^2$-Lab, Mediterranean Center of Human Health and Advanced Biotechnologies \\ University of Palermo, Viale delle Science, Edificio 8, 90100 Palermo, Italy}

\address[address6]{Carnegie Mellon University, Pittsburgh PA 15213-3890, USA}

\begin{abstract}
In this work lipid ordering phase changes arising in planar membrane bilayers is investigated both accounting for elasticity alone and for  effective viscoelastic response of such assemblies. The mechanical response of such membranes is studied by minimizing the Gibbs free energy which penalizes perturbations of the changes of areal stretch and their gradients only \cite{Deseri:2013}. As material instabilities arise whenever areal stretches characterizing homogeneous configurations lie inside the spinoidal zone of the free energy density, bifurcations from such configurations are shown to occur as oscillatory perturbations of the in-plane displacement.  Experimental observations \cite{Espinosa:2011} show a power-law in-plane viscous behavior of lipid structures allowing for an effective viscoelastic behavior of lipid membranes \cite{DeseriZingales:2015}, which falls in the framework of  Fractional Hereditariness. A suitable generalization of the variational principle invoked for the elasticity is applied in this case, and the corresponding Euler-Lagrange equation is found together with a set of boundary and initial conditions.
Separation of variables allows for showing how Fractional Hereditariness owes bifurcated modes with a larger number of spatial oscillations than the corresponding elastic analog. Indeed, the available range of areal stresses for material instabilities is found to increase with respect to the purely elastic case. Nevertheless, the time evolution of the perturbations solving the Euler-Lagrange equation above exhibits  time-decay and the large number of spatial oscillation slowly relaxes, thereby keeping the features of a long-tail type time-response.
\end{abstract}


\end{frontmatter}

\section{Introduction}
Lipid bilayers are known to be building blocks of almost all types of biological membranes, as they surround the cells of almost of all living organisms. In the last decade, the growing availability of advanced microscopy and imaging techniques has determined a blooming of interest in the study of biological membranes, often revealing  spectacular examples of intricate patterns at micro and nano scales (see, {\it e.g.}, \cite{Baumgart:2003}). 

The intimate presence of lipids in the cell membrane strongly influences its multiphysics and, hence, its mechanical behavior. Of course this is highly dependent on a rich list of parameters such as the configuration assumed by the lipids, the chemical composition, temperature of their watery environment and applied osmotic pressure \cite{Bermudez:2004, Das:2008, Iglic:2012, Sackmann:1995, Hu:2012, Norouzi:2006, Agrawal:2008, Agrawal:2009, Agrawal:2015, Baumgart:2005}.

In particular, these amazing structures are capable to sustain bending moments and normal stress, due to their special constitutive nature, showing ordering-disordering phenomena which allow changes in the shape for responding to the external solicitations. The pioneering works on modeling the the bending behavior of biological membranes can be traced back to Canham \cite{Canham:1970} and Helfrich \cite{Helfrich:1973}. These models relie upon the assumptions of (i) ``in-plane fluidity'' and (ii) elasticity of the membrane, hence in-plane shear stress cannot arise.

Other studies on the equilibrium shapes of biomembranes include the influence of presence of embedded proteins \cite{Canham:1970, Jenkins:1977, Agrawal:2009, Biscari:2002}.

The ordering-disordering phenomena have been extensively investigated \cite{Akimov:2003, Chen:2001, Falkovitz:1982, Goldstein:1989, Iglic:2012, Jahnig:1981, Owicki:1978, Owicki:1979} in order to understand their influence on the mechanical behavior of the biological membranes. This leads to the formation of buds  \cite{Lipowsky:1992}, but this transition can be also related to the molecules structure \cite{Komura:2004, Pan:2009, Rawicz:2000}. 

The energetics governing the thermo-chemo-mechanical behavior of this structures was recently derived \cite{Deseri:2008, Deseri:2013, Agrawal:2009, Fried:2013} for a better understanding of the mechanics of the biological membranes and a powerful tool for predicting their response whenever specific conditions occur. 

The main feature of this approach is that the energetics of the membrane can be described through one single ingredient: the in-plane membrane stretching elasticity. This allows for describing the response with respect to local area changes on the membrane mid-surface. The principle of the minimum of energy allows for characterizing the governing equation of the mechanical response of the membrane. This approach allows for determining the profile and the boundary layer of the disordering-ordering phenomena, i.e the change from a thicker domain (ordered phase) to a thinner one (disordered phase), and their associated rigidities.

The main feature of the energy derived in \cite{Deseri:2010} is the presence of two turning points in the local stress governing the biological membrane behavior (see \Fig{fig:fig_dz_energy}a). They are placed in a region characterized by material instabilities, i.e in a spinoidal zone. Henceforth, whenever the external conditions are such that the areal stretch, i.e. the reciprocal of the thinning, is enclosed in this region, the response may produce a  rapid change of the geometry, i.e material instabilities can occur. In this work, we show that this occurrence is exhibited even when the in-plane viscosity of the lipid membrane is accounted for. In this regard, the experimental observations of lipid viscous behavior showed that the loss and storage moduli are well described by power law functions \cite{Espinosa:2011}. This observation suggests that the viscoelastic behavior of the biological membrane is properly described in the framework of the Fractional Hereditariness. Indeed, upon introducing an enriched kinematics accounting for in-plane shears and the exhibited in-plane power-law viscosity in a parallel contribution \cite{DeseriZingales:2015}, a dimension reduction procedure analog to one shown in \cite{Deseri:2008, Deseri:2013} will be used for studying the fractional viscoelastic behavior mentioned above.

The onset of bifurcated configurations possibly arising from homogeneous configurations characterized by an areal stretch lying in the spinoidal region is studied in Section \ref{chap:elastic0}. Here we minimize the total elastic (Gibbs free) energy to determine the bifurcated modes and the relationships  between the number of nucleated spatial waves with the critical values of the areal stretches.

The influence of the effective viscoelasticity on the material instabilities exhibited by the membrane is studied in Section \ref{chap:viscous}.\\
\newline 
The problem is formulated by seeking for the values of the areal stretches for which unknown time evolving bifurcated configurations could occur. To this aim, in full analogy with the elastic case, a variational principle is employed. Here, the Gibbs free energy density is taken from \cite{Deseri:2014}, where a rheological model yields the Staverman-Schartzl free energy \cite{DelDes:1996, DelDes:1997,DesGoldFab:2006, DesGold:2007} as the one for power-law materials.

As in the elastic case, the viscoelastic free energy has a local and a nonlocal part. There, the power at which stress and hyperstress (which performs work against changes of the displacement gradient $u_x$, see \cite{Deseri:2013} for more details) relax could be different, as diffusion mechanisms may occur at different average speed depending on whether or nor they arise in a boundary layer between different phase or in a given phase.

\section{The membrane elasticity theory for the lipid bilayers}
\label{chap:elastic0}
In this section we briefly recall the main results obtained in \cite{Deseri:2008, Zurlo:2006}, together with a schematic description of the approach followed in the papers. There the formulation of the membrane problem is restricted to initially planar membranes, i.e. the effects of spontaneous curvature have been neglected. In this case a simplified version of the elastic energy for the configuration change of the membrane geometry is obtained.


An orthonormal reference frame $(\be_1,\be_2,\be_3)$ is introduced and \black{a prismatic region $\Body_0$ of constant thickness $h_0$} is taken as reference configuration. A flat mid-surface $\Omega$ in the plane spanned by $(\be_1,\be_2)$ is singled out for further use. Points of $\Body_0$ are denoted by
\begin{equation}
\bx=\sx+z\be_3,
\end{equation}
where $\sx=x\,\be_1 + y\,\be_2$ and $z\in(-h_0/2,h_0/2)$. Denote by $\ff$ the deformation map and by $\FF=\nabla\ff$ its gradient. Thus, the stored Helmholtz free-energy can be expressed as
\begin{equation}
\scrE(\ff)=\int_{\Body_0}W(\FF)\,dV=\int_{\Omega} \int_{-h_0/2}^{h_0/2}W(\FF)\,dz\,d\Omega,
\end{equation}
where $W$ is the purely elastic Hemholtz energy density per unit \black{volume}. The surface energy density is, then,
\begin{equation}
\label{eq:en2d}
\psi(\ff)=\int_{-h_0/2}^{h_0/2}W(\FF)\,dz.
\end{equation}
Lipid membranes are known to be characterized by {\it in-plane fluidity}, corresponding to the impossibility of sustaining shear stresses in planes perpendicular to $\be_3$, unless some viscosity is present. This constitutive assumption can be used to restrict the pointwise dependence $W$ on a list of three invariants of $\FF$ (see \cite{Deseri:wip:1} for details)
\begin{equation}
\scrI(\bx)=\{\tilde{J}(\bx),\,\det\FF(\bx),\,\bar{\phi}(\bx)\},
\end{equation}
which can be interpreted as the areal stretch of planes perpendicular to the direction $\be_3$, the volume variation and the stretch in direction $\be_3$, i.e. the thickness stretch $\bar{\phi}(\bx) = h(\bx)/h_0$, respectively.
%
\begin{figure}[h]
\centering
\includegraphics[width=8cm]{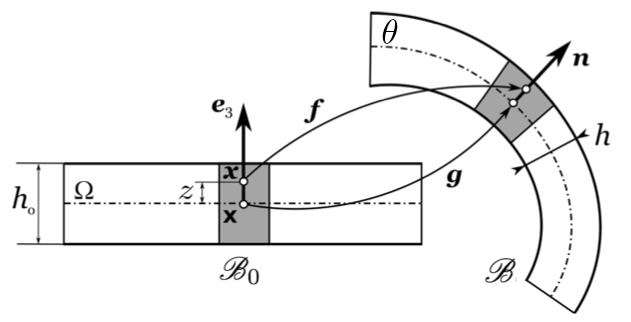}
\caption{Schematic representation of the deformation (\ref{eq:def}) of a prismatic, plate-like reference configuration $\Body_0$ into the current configuration $\Body$. The gray box depicts the space occupied by two lipid molecules, their volume being conserved during the deformation. Courtesy of \cite{Deseri:2013}}
\label{fig:fig00_schemagen}
\end{figure}

In order to capture the out-of-plane deformations of the membrane and the occurrence of  thickness changes, the following ansatz (see Fig.\ref{fig:fig00_schemagen}) has been assumed
\begin{equation}
\label{eq:def} 
\ff(\bx) = \gg (\sx) + z\phi(\sx)\,\bn(\sx),
\end{equation}
where $\gg(\sx)=\gg(x,y,0)$ defines the current mid-surface of the membrane, that is $\theta=\gg(\Omega)$, where $\bn$ is the outward normal to $\theta$ and where $\phi(\sx)=h(\sx)/h_0$ is the thickness stretch, with  $h$ the current thickness. Such ansatz permits to make explicit the dependence of the invariants $\scrI$ on $z$ and, ultimately, to perform the expansion of (\ref{eq:en2d}) in powers of the reference thickness $h_0$.

The molecular volume of biological membranes can be considered almost constant in a wide range of temperature \cite{Goldstein:1989, Owicki:1978}. Because \eqref{eq:def} holds, this condition can be imposed by means of a {\it quasi - incompressibility} constraint
\begin{equation}
\label{eq:quasinc}
\det\FF(\sx,0)=\tilde{J}(\sx,0) \phi(\sx) = 1.
\end{equation}

The constraint (\ref{eq:quasinc}) is first order approximation of the exact incompressibility requirement, since $\det\FF(\bx)=\det\FF(\sx,0) + O(z)$ for a planar deformations, the condition (\ref{eq:quasinc}) implies that $\det\FF(\bx)=1$ holds exactly. This is the special case considered in this work.

It is then appropriate to introduce the restriction of the Helmholtz energy density $W$ to $\Omega$ for quasi-incompressible deformations, 
\begin{equation}
w(J)=W(\tilde{J},\det\FF,\bar{\phi}){\Big |}_{z=0}=W(J,1,J^{-1}) ,
\end{equation}
where $J(\sx)=\tilde{J}(\sx,0)$.

At this point, under ansatz (\ref{eq:def}) and the assumptions of in-plane fluidity and bulk incompressibility, the expansion of (\ref{eq:en2d}) up to $h_0^3$ gives 
\begin{equation}
\label{eq:en2dexp}
\psi = \varphi(J) + \kappa(J) H^2 + \kappa_G(J) K + \alpha(J) \, ||\left(\grad_{\red{\theta}}\wh J \right)_{\scriptscriptstyle m}||^2 ,
\end{equation}
where  $H$ and $K$ are the mean and Gaussian curvatures of the mid-surface $\theta$, respectively, $\kappa(J)$ and $\kappa_G$ are the corresponding bending rigidities and 
\begin{equation}
\label{eq:alpha}
\alpha(J)=\frac{h_0^2}{24}\frac{\f'(J)}{J^5}.
\end{equation}
In equation \eqref{eq:en2dexp}, $\hat{J}$ is the spatial description of $J$ , defined by the composition $\hat{J} \circ g = J$, $\grad_{\theta}$ is the gradient with respect to points of the current mid-surface $\theta$, and $(\cdot)_{\scriptscriptstyle m}  $ gives his material description.

The main ingredient of the two-dimensional membrane model derived in (\ref{eq:en2dexp}) is the surface Helmholtz energy $\f(J)$, which regulates the in-plane stretching behavior of the membrane and describes the phase transition phenomena taking place in lipid bilayers. In fact, due to increase in temperature  the (hydro)carbon tails  of phospholipid molecules undergo a (first-order) phase transition, i.e. a thickness reduction from the liquid ordered phase $L_o$ to the liquid disordered phase $L_d$. Due to the constraint $J\phi=1$, both $J$ and $\phi$ have been adopted in literature as coarse-grained order parameters for the study of the $L_o-L_d$ transition (see, {\it e.g.}, \cite{Falkovitz:1982, Goldstein:1989, Jahnig:1981, Jahnig:1996, Owicki:1978, Owicki:1979, Sackmann:1995}).
%
\begin{figure}[ht]
\centering
\includegraphics[width=0.9\columnwidth]{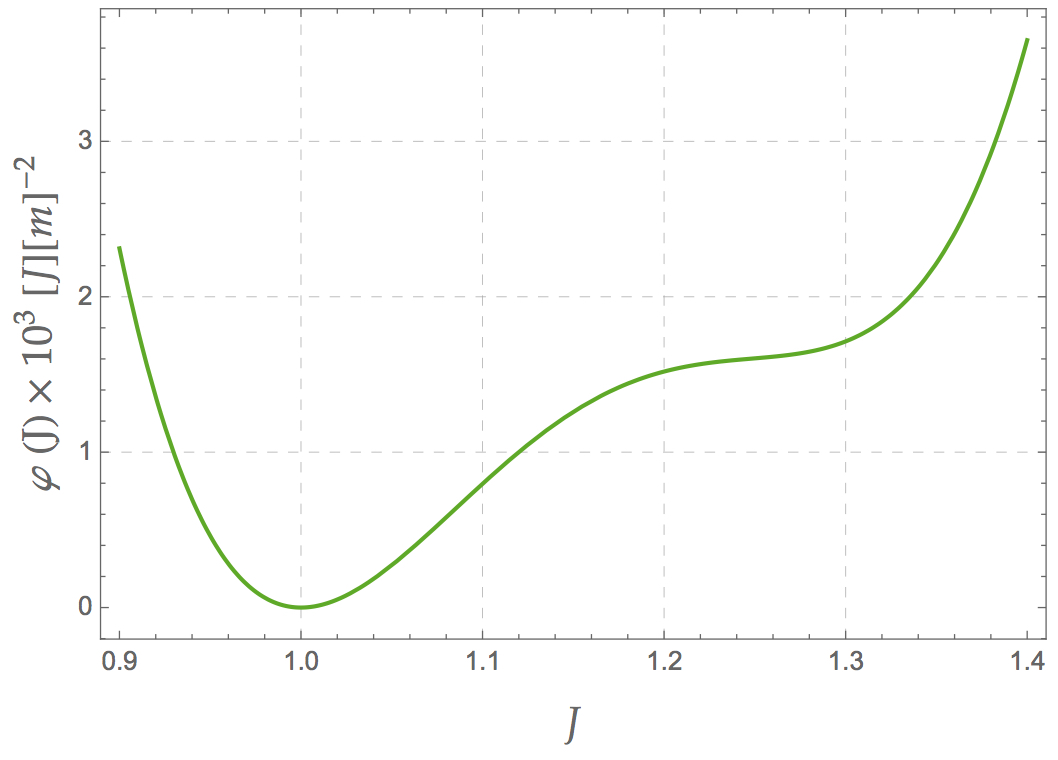} \quad
\includegraphics[width=0.9\columnwidth]{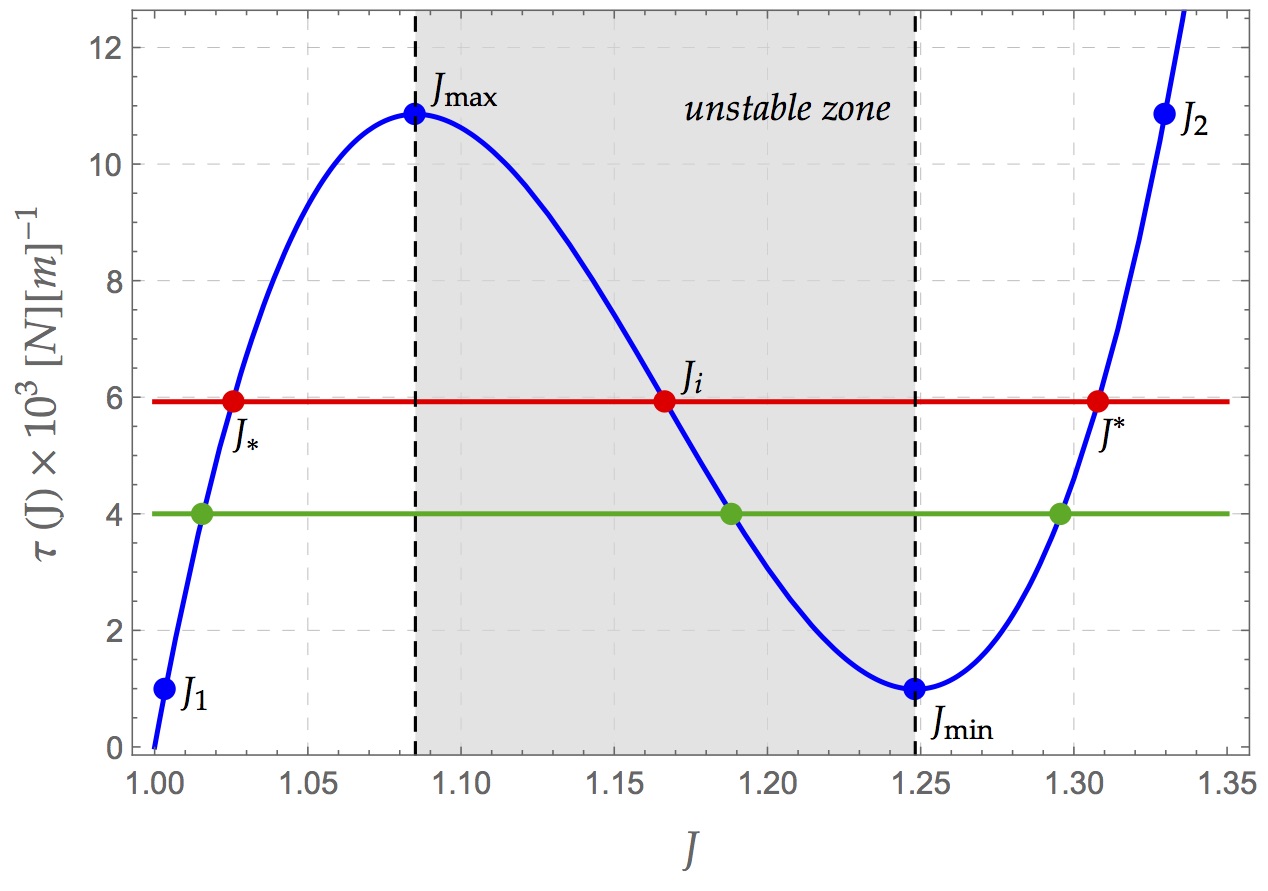} 
\caption{The stretching energy $\f(J)$ adapted from \cite{Goldstein:1989} for a temperature $T\sim 30^{\circ}$ and related local stress $\f'(J) = \tau(J)$. The areal stretch $J_o=1$ corresponds to the unstressed, reference configuration $\Body_0$. Courtesy of \cite{Deseri:2013}}
\label{fig:fig_dz_energy}
\end{figure}

Experimental evidence clearly shows that for a given chemical composition there exists a temperature range where the $L_o$ and $L_d$ phases coexist, organizing themselves in domains called {\it rafts}.

A classical method to determine $\f(J)$ is the construction of an appropriate  Landau expansion of the stretching free energy in powers of the order parameter $J$ (see, {\it e.g.}, \cite{Falkovitz:1982, Goldstein:1989, Komura:2004, Owicki:1978, Owicki:1979}). The advantage of the Landau expansion is that its parameters can be related to measurable quantities, such as the transition temperature, the latent heat and the order parameter jump (see \cite{Goldstein:1989} and the treatise \cite{Sackmann:1995} for a detailed discussion).

By assuming that for a fixed temperature the membrane natural configuration $\Body_0$ coincides with the flat, ordered $L_o$ phase, in which $J=J_o=1$, the stretching energy is chosen in the form
\begin{equation}
\label{eq:phi}
\f(J) = a_0 + a_1 J + a_2 J^2 + a_3 J^3 +a_4 J^4,
\end{equation}
where the parameters $a_i\,\,(i=0,...,4)$ depend in general on temperature and chemical composition. In the lack of specific experimental data and in order to show the numerical feasibility of the model, we calibrate these parameters on the basis of the experimental estimates provided by \cite{Goldstein:1989, Komura:2004, Komura:2006}. For a temperature $T\sim 30^{\circ}$, we have
\begin{equation}
\label{eq:ai}
\begin{array}{c}
a_0=2.03 , \quad a_1=-7.1, \quad a_2=9.23 \\
a_3=-5.3, \quad a_4=1.13,
\end{array} 
\end{equation}
dimensionally expressed in $[J][m]^{-2}$. It is worth pointing out that this specific choice is illustrative and is meant to show the feasibility of the current approach.

\subsection{Planar case}
The study of the equilibrium problem for a planar lipid membrane described by the energy (\ref{eq:phi}) with the constants given by (\ref{eq:ai}) permits to elucidate the emergence of thickness inhomogeneities in the membrane and allows one to calculate the corresponding rigidities and the shape of the boundary layer between the ordered and disordered phases. Whenever no curvature changes are experienced by the lipid bilayer, the elastic energy density in \eqref{eq:en2dexp} takes the form:
\begin{equation}
\label{eq:energy}
\psi_{DZ}=\varphi\left(J\right)+\alpha(J) ||\grad_{\theta}\wh J||^2.
\end{equation}
In this work, following \cite{Coleman:1988}, we consider a membrane that in the reference configuration $\Body_0$ has the form of a thin plate of homogeneous thickness $h_0$ (direction $\be_3$), width $B$ (direction $\be_2$) and length $L$ (direction $\be_1$). The reference membrane mid-surface $\theta$ corresponds to $z=0$, and its edges are defined by $x=\pm L/2$ and $y=\pm B/2$.  Henceforth, the three-dimensional membrane deformation is further restricted with respect to (\ref{eq:def}), according to
\begin{equation}
\label{eq:def2}
\ff(\bx)=g(x)\be_1+y\be_2+z\phi(x)\be_3
\end{equation}
so that the width $B$ is kept constant and its gradient takes the following form
\begin{equation}
\label{eq:F}
\FF=\nabla\ff=
  \left[
   \begin{array}{ccc}
      g_{x} & 0 & 0 \\
      0 & 1 & 0 \\
      z\phi_{x} & 0 & \phi
   \end{array}
   \right],
\end{equation}
where the subscript $x$ denotes differentiation with respect to $x$. The displacement component along $\be_1$ is $u(x)=g(x) - x$. After setting
\begin{equation}
\label{eq:lambda}
\lambda(x)= g_{x}(x)
\end{equation}
for the stretch in direction $\be_1$, we have $\det\FF=\lambda\phi = 1$ and $J=\lambda$. Hence $\phi=\lambda^{-1}$, so that the membrane deformation is completely determined by $J = \lambda$. In \cite{Deseri:2013}, the Euler-Lagrange equation related to this kinematics and the same form of local energy  \eqref{eq:phi} was deeply studied, obtaining the following result:
\begin{equation}
\label{eq:equilibrium_dz}
\gamma(J)\, J + \frac{1}{2} \gamma(J) \, J_x^2 + \tau(J) = \Sigma
\end{equation}
where $\gamma(J) = 2 \alpha(J)$ and $\Sigma$ is a force per reference length on the edges $x = \pm L/2$. In such conditions, it is easy to show that homogeneous configurations are in the set of equilibria. Indeed, whenever an homogeneous configuration is considered, the higher-order terms drop to zero and the equilibrium equation reads as:
\begin{equation}
\tau(J) = \Sigma
\end{equation}
The special form of the local stress $\tau(J) = \f'(J)$ shown in \Fig{fig:fig_dz_energy} allows for discriminating several cases around the spinodial-zone, i.e. where the function $\tau(J)$ is an S-shaped function. Indeed, whenever $J < J_1$ and $J>J_2$ the equilibrium can be reached for only one value of $J$, namely $\Sigma = \tau(J)$. On the contrary, if $J_1 < J < J_2$ the configuration lies in the spinoidal-zone, and the membrane can sustain the same value of the stress by assuming three different configurations, i.e. the three intersection of the function $\tau(J)$ with the horizontal straight line representing the values of the stress at the edges. Here, the only parameter governing the membrane behavior is the areal-stretch $J$, henceforth, by recalling the basic idea of the instabilities of structures, the system is stable if the second derivative of the total potential energy (namely $\f(J)$ for an homogeneous configuration) is positive. Therefore, two different behaviors occur inside the spinoidal zone: if $J_1<J<J_{max}$ or $J_{min} < J < J_2$ the second derivative of the energy is positive $\f''(J) > 0$ (i.e, the slope of $\tau(J) = \f'(J)$ is positive), and the behavior is stable, otherwise $J_{max}<J<J_{min}$ and the second derivative assumes negative values, namely $\f''(J) < 0$  and the slope of $\tau(J) = \f'(J)$ is negative,  determining the unstable behavior.
The only interesting phenomena due to a perturbed configuration arise whenever the membrane, for some reasons (e.g. a configuration imposed in a experimental setup), is homogeneously stretched with a value lying in the unstable zone.

\subsection{The linearized mechanics of membrane elasticity}
In this section we obtain the linearized equation of lipid membrane under the plane strain geometry \eqref{eq:F} with $g_x = \bar{J}$ and $\phi = \bar{\phi}$ (hence $\phi_x=0$). In this regard let us denote with $\e$ the strain field perturbing uniformly the stretched configuration  just described. The elastic free energy density \eqref{eq:en2dexp} for the membrane is then evaluated at the perturbed configuration $J=\bar{J}+\e$, and takes the form:
\begin{equation}
\label{eq:e_energy_density}
\begin{aligned}
\psidz \left(\e, \,\e_x \right) & = \f\left(\bar{J}+\e\right)+\alpha(\bar{J}+\e) ||\left( \bar{J} + \e\right)_x||^2 \\
& \approx \varphi(\bar{J})+\f'(\bar{J})\,\e+\frac{\f''(\bar{J})}{2} \e^2+\alpha(\bar{J}) \,||\e_x||^2
\end{aligned}
\end{equation}
where we neglected higher-order contributions in $\e^2$.  Then the free energy takes the form:
\begin{equation}
\label{eq:energy_dz}
\Psi_{DZ}=\int_{\Omega}\psidz(\e,\e_x)dx,
\end{equation}
where a domain $\Omega \in [-L/2,L/2]$ is considered and
\begin{equation}
\label{eq:quadratic_energy}
\psidz(\e,\e_x) = \varphi(\bar{J})+\f'(\bar{J})\,\e+\frac{\f''(\bar{J})}{2} \e^2+\alpha(\bar{J}) \, \e_x^2.
\end{equation}
As consequence of this choice, the (in-plane) displacement field is described through a perturbation $v$ such that $u = \bar{u} + v$. Of course, $\e(x) = v_x(x)$.

We assume that the membrane is pulled by opposite tractions of magnitude $\Sigma$ (force per reference length) at the boundary, i.e on the edges $x=\pm L/2$, although the case in which the end displacements are controlled may be treated in an analog way (see, {\it e.g.}, \cite{Triantafyllidis:1993}). Due to the presence of nonlocal terms $\e_{x}$, it is necessary to introduce \emph{hyper-tractions} $\Gamma$ which perform work against displacement gradient $v_x$ at the boundary \cite{Puglisi:2007}. Henceforth, the total energy $\scrE$ change in a neighborhood of the homogeneously deformed configuration reads as follows:
\begin{equation}
\label{eq:totpoten}
\scrE=B \,\Psidz - \scrW(v,v_x),
\end{equation}
where $B$ denotes the width of the membrane patch and $\scrW$ is the external work of the applied tractions $\Sigma$ and hypertractions $\Gamma$ (see \cite{Deseri:2013}) defined as follows:
\begin{equation}
\label{eq:e_work}
\scrW(v,v_x) = B \dO{\Sigma \,(\bar{u} + v) +\Gamma \,(\bar{u}_x +  v_x)} ,
\end{equation}
where $\bar{u} = \bar{J}_x$ is the displacement corresponding to the homogeneously  stretched configuration from which bifurcations are sought. Upon substituting \eqref{eq:e_energy_density} and \eqref{eq:totpoten} in \eqref{eq:e_work}the total energy change takes the following form:
\begin{equation}
\label{eq:en_functional}
\begin{aligned}
\scrE & = B \intO \left(\varphi +\f'(\bar{J})\,v_x+\frac{\f''(\bar{J})}{2} v_{x}^2+\alpha(\bar{J}) \, v_{xx}^2 \right) dx  \\
&- B \dO{\Sigma \, v +\Gamma \, v_x} + \bar{\scrE} .
\end{aligned}
\end{equation}
The variation of the energy is computed with respect to a reference value $\scrE (\bar{J})$ defined as follows:
\begin{equation}
\label{eq:ref_energy}
\bar{\scrE} = B \intO \f(\bar{J}) dx-  \dO{\Sigma \, \bar{u}+\Gamma \, \bar{u}_x}.
\end{equation}
In the sequel all the quantities with the over-bar are referred to the homogeneously stretched configuration, e.g. $\bF := \f(\bar{J})$,  $\bF'' := \f''(\bar{J})$ and $\bA := \alpha(\bar{J})$. 

The resulting governing equation of the planar membrane is obtained by imposing the stationarity of $\scrE$ (see \ref{chap_A1} for details). Such equation together with its boundary conditions reads as follows:
\begin{equation}
\label{eq:euler_elastic_1}
\left\{
\begin{array}{ll}
2 \bA \, v'''' - \bF'' \, v'' = 0   & \text{in} ~\Omega\\ 
\text{either}~ \bF'' \, v' - 2\bA \, v''' = \Sigma - \bF ~\text{or}~ \d v = 0 & \text{in}~ \partial \Omega  \\
\text{either}~ 2 \bA \, v'' = \Gamma  ~\text{or}~ \d v' = 0 & \text{in}~ \partial \Omega  
\end{array} 
\right.
\end{equation}
It is worth noting that homogeneous configurations of the membranes from which oscillatory perturbations could arise are not known. In order to find the values of $\bar{J}$ characterizing such homogeneous states and to study the solution of the boundary value problem governing bifurcated equilibria from  such configurations, a parameter $\omega$ is introduced as follows:
\renewcommand*{\arraystretch}{2}
\begin{equation}
\label{eq:omega}
\omega^2 := \left\{
\begin{array}{ll}
+ \dfrac{\bF''}{2 \bA}   \quad \text{if } \bF'' > 0 \\
- \dfrac{\bF''}{2 \bA}   \quad\text{if } \bF'' < 0 ,
\end{array} 
\right.
\end{equation}
\renewcommand*{\arraystretch}{1}
where:
\begin{equation}
\label{eq:e_f2_f1_1}
\frac{\bF''}{2 \bA} = \frac{12}{h_0^2} \frac{\bF''}{\bF'} \bar{J}^5,
\end{equation}
because of \eqref{eq:alpha}. Henceforth, equation \eqref{eq:euler_elastic_1} can be recast as:
\renewcommand*{\arraystretch}{1.2}
\begin{equation}
\label{eq:euler_elastic_2}
\left\{
\begin{array}{ll}
v'''' \mp \w^2\, v'' = 0   & \text{in} ~\Omega\\ 
\text{either}~ \pm \w^2 v' - \, v''' = \dfrac{\Sigma - \bF}{2\bA} ~\text{or}~ \d v = 0 & \text{in}~ \partial \Omega  \\
\text{either}~ 2 \bA \, v'' = \Gamma  ~\text{or}~ \d v' = 0 & \text{in}~ \partial \Omega   .
\end{array} 
\right.
\end{equation}
\renewcommand*{\arraystretch}{1}
The choice of the boundary conditions above generate various cases.  For the sake of illustration, we choose the case in which the displacement is constrained and the hypertractions are imposed at the boundary, i.e. $v = 0$ and $2 \bA \, v'' = \Gamma$.

It is worth noting that the assumed value of $\omega^2$ affects the quality of the solution, i.e. the onset of phase changes in the elastic membrane. In this regard some sub-cases can be identified depending upon the location of the reference condition associated to $\bar{J}$ in the stretching energy function in \Fig{fig:fig_dz_energy}.  Indeed, because $\varphi(J)$ has at most one stationary point $J_0$ unless the lipid bilayer is at its transition temperature,  inspection of \Fig{fig:fig_dz_energy} shows that there are four values of $J$ besides $\bar{J}$ to be accounted for, namely $J_{*}\leq J_{max}\leq J_{min}\leq J^{*}$. Here $J_{max}, \, J_{min}$ are stationarity points of $\tau(J) = \varphi'(J)$, i.e. $\varphi(J)$ changes curvature there (namely $\f''(J)$ changes sign, while $J_{*}$ and $J^{*}$ are the abscissas of the two points sharing common tangent on $\varphi(J)$. Two alternative situations may arise depending on the sign of $\bF''$. This depends on whether or not the configuration $\bar{J}$ is in the spinoidal (\textit{unstable}) zone of the local energy density $\f(J)$.

\subsection{Unstable zone: $\bar{\varphi}{''}< 0$}
We explore the case for which $\bar{\varphi}{''}< 0$ in \eqref{eq:omega}, which happens whenever $\bar{J}$ is located in the spinoidal zone, i.e. $J_{max} < \bar{J} < J_{min}$, corresponding to a negative slope of the local stress,  since $\tau(J) = \f'(J)$ (see \Fig{fig:fig_dz_energy}). The governing equation \eqref{eq:euler_elastic_2} takes the following form: 
\begin{equation}
\label{eq:e_f<0}
v_{xxxx}+ \omega^2 v_{xx}=0 ,
\end{equation}
which admits the integral
\begin{equation}
\label{eq:e_f<0_sol}
v(x) = A_1 \, \cos (\w \, x) + A_2 \sin(\w \, x) + A_3 \, x + A_4.
\end{equation}
We explore this solution for the following boundary conditions apply:
\begin{equation}
\label{eq:BC_el}
v\Big|_{\partial \Omega^-} = 0 \qquad v\Big|_{\partial \Omega^+} = 0 \qquad 2\bA v'' \Big|_{\partial \Omega^-} = \hat{\Gamma}_{\sss{L}} \qquad 2\bA v''\Big|_{\partial \Omega^+} = \hat{\Gamma}_{\sss{R}} 
\end{equation}
where $\hat{\Gamma}_{\sss{R}}  = \Gamma \Big |_{\pd \Omega^{+}}$ and $\hat{\Gamma}_{\sss{L}}  = \Gamma \Big |_{\pd \Omega^{-}}$. The values of the coefficients $A_i$ in \eqref{eq:e_f<0_sol} depend on the specified boundary conditions. For the sake of convenience the positions $c = \cos(\w \, L/2)$ and $s = \sin(\w \, L/2)$ are assumed; henceforth, the BCs assume the following form:
\begin{equation*}
\begin{aligned}
&
\left\{
\begin{array}{l}
A_1 \, c - A_2 \,s - A_3 \dfrac{L}{2} + A_4 = 0 \\
2 \bA \w^2 \left(-A_1 \, c + A_2 \, s \right) = \hat{\Gamma}_{\sss{L}}
\end{array}
\right.
\,\text{at}\, x = -\frac{L}{2} \\
&
\left\{
\begin{array}{l}
A_1 \, c + A_2 \,s + A_3 \dfrac{L}{2} + A_4 = 0 \\
2 \bA \w^2 \left(-A_1 \, c - A_2 \, s \right) = \hat{\Gamma}_{\sss{R}}
\end{array}
\right.
\,\text{at}\, x = +\frac{L}{2} \\
\end{aligned}
\end{equation*}
%
In this example we assume $\hat{\Gamma}_{\sss{L}} = \hat{\Gamma}_{\sss{R}} = \hat{\Gamma}$. These assumptions lead to a simplified matrix system:
\renewcommand*{\arraystretch}{1.25}
\begin{equation}
\label{eq:sys_el}
\left[
\begin{array}{cccc}
0 & s & \frac{L}{2} & 0 \\ 
c & 0 & 0 & 1 \\ 
0 & s & 0 & 0 \\ 
-2\bA \, \w^2 c & 0 & 0 & 0
\end{array} 
\right]
 \, 
\left(
\begin{array}{c}
A_1 \\ A_2 \\ A_3 \\ A_4
\end{array} 
\right)
=
\left(
\begin{array}{c}
0 \\ 0 \\ 0 \\ \hat{\Gamma}
\end{array} 
\right) ,
\end{equation}
whose determinant is $\bA \, c \, s \, L \, \w^2$. We first study the nontrivial modes \eqref{eq:e_f<0_sol} of the system, i.e. we explore the roots of the following equation
\begin{equation}
\bA \, c \, s \, L \, \w^2 = 0 .
\end{equation}
It is worth noting that, because of the definition \eqref{eq:alpha} and $1 < J_{max} < \bar{J} < J_{min}$, we have $\bA > 0 $ for all  $\bar{J}>1$. Then, the orthogonality of the trigonometric functions imposes that the equation is satisfied if either for $c = \cos(\w \, L/2) = 0$ or for  $s = \sin(\w \, L/2) = 0$. Henceforth, we are left to examine two subcases.
\quad \\
\newline
\textbf{Case 1.} Let us consider the case $s = 0$ and $c = \pm 1$. This condition implies that:
\begin{equation}
\label{eq:e_n_case1}
\sin \left(\w \, 	\frac{L}{2}\right) = 0 \quad \Longrightarrow \quad \w \, \frac{L}{2} = n \, \pi \quad \Longrightarrow \quad  \w = \frac{2\, n \, \pi}{L}
\end{equation}
and a closer analysis of \eqref{eq:e_n_case1} shows that this case occurs whenever the following relationship holds:
\begin{equation}
\label{eq:relationship1}
\frac{\bF''}{\bF'} \bar{J}^5 = -\frac{n^2 \pi^2}{3} \left(\frac{h_0}{L} \right)^2 .
\end{equation}
The thinness of the membrane here enters with the ratio $\left(h_0/L \right)^2$ which is normally smaller than $10^{-8}$. A large but finite number $n$ of oscillation can certainly arise from \eqref{eq:relationship1} for $J$ such that $\bF'' \to 0^-$, i.e. right after \black{change on convexity} of the local part  of the strain energy density. The solution of the system allows for deducing the values of amplitude of the $n^{th}$ mode:
\renewcommand*{\arraystretch}{1.25}
\begin{equation*}
\left[
\begin{array}{cccc}
0 & 0 & \frac{L}{2} & 0 \\ 
\pm 1 & 0 & 0 & 1 \\ 
0 & 0 & 0 & 0 \\ 
\mp 2\bA \, \w^2 & 0 & 0 & 0
\end{array} 
\right]
 \, 
\left(
\begin{array}{c}
A_1 \\ A_2 \\ A_3 \\ A_4
\end{array} 
\right)
=
\left(
\begin{array}{c}
0 \\ 0 \\ 0 \\ \hat{\Gamma}
\end{array} 
\right)
\end{equation*}
then
\begin{equation*}
\left\{
\begin{array}{l}
A_1 = \mp \dfrac{\hat{\Gamma}}{2\bA \, \w^2} \\
A_3 = 0 \\
A_4 = \mp A_1
\end{array}
\right. .
\end{equation*}
Hence, the buckled mode $n$ has the following form:
\begin{equation}
\label{eq:e_modes}
\begin{aligned}
v_n(x)  & = \pm \dfrac{\hat{\Gamma}}{8 \,\bA \,n^2 \,\pi^2 } \left[ \cos \left(2n\pi \frac{x}{L}\right)  - 1\right] \\
& \quad + A_2 \sin \left(2n\pi \frac{x}{L}\right) .
\end{aligned}
\end{equation}
It is worth noting that even if the hyperstress $\hat{\Gamma}$ at the boundary would vanish, equation \eqref{eq:e_modes} assures that a bifurcation always occurs with a bifurcated mode $v_n=A_2 \sin \left(2 n \pi\frac{x}{L}\right)$. 

It is natural to ask if there is a reduction of energy by nucleating oscillations.
The amount of the extra energy for getting the final configuration from $\bar{J}$ is computed in \ref{chap_A2}. It turns out that it is identically zero. This fact suggests that all the buckled configuration from $\bar{J}$ posses the same quantity of energy and then such buckled configuration do have the same likelihood to occur.
%
\quad \\
\newline
\textbf{Case 2.} Let us now consider the case $s = \pm 1$ and $c = 0$. This condition implies that:
\begin{equation}
\cos \left(\w \, 	\frac{L}{2}\right) = 0  \quad \Longrightarrow \quad  \w = \frac{(1+2\, n) \, \pi}{L}
\end{equation}
and
\begin{equation}
\label{eq:e_f2_f1_2}
\frac{\bF''}{\bF'} \bar{J}^5 = -\frac{(1+2n)^2 \pi^2}{12} \left(\frac{h_0}{L} \right)^2,
\end{equation}
which has certainly roots for $\bar{J}$ such that $\bF'' \to 0^-$ for the reason explained in case 1. As usual, the coefficients of the mode are found by imposing the boundary conditions. We find  $A_2 = A_3 = A_4 = 0$. Hence, in this case a solution is possible if and only if $\hat{\Gamma} = 0$. It follows that the buckled modes take the forms:
\begin{equation}\
\label{eq:e_f>0_sol}
v_n(x) = A_1 \, \cos(\w \, x)  = A_1 \, \cos \left((1+2n)\pi \frac{x}{L}\right).
\end{equation}
It is easy to recognize that also in this case the extra amount of energy needed to bifurcate from $\bar{J}$ is equal to 0.

\subsection{Stable zone: $\bar{\varphi}{''}> 0$}
Whenever the configuration of the membrane $\bar{J}$ is located outside of the spinoidal zone, i.e $\bF'' > 0$ and  either $1<\bar{J} < J_{max}$ or $\bar{J} > J_{min}$, the governing equation assumes the following form:
\begin{equation}
v'''' - \w^2 \, v'' = 0.
\end{equation}
In such a case, the profile of the perturbation becomes: 
\begin{equation}
v(x) = A_1 \, \cosh(\w \, x) + A_2 \, \sinh(\w \, x)  + A_3 \, x + A_4,
\end{equation}
where the coefficients $A_i$, as in the previous analysis, depend of the specific boundary conditions.

\subsection{Singular points: $\bar{\varphi}{''} = 0$}
Before proceeding further some additional discussion may be withdrawn from the analysis of the singular points $\bar{J}=J_{max}$ and $\bar{J}=J_{min}$. In both cases, the first derivative of the local stress is zero, i.e $\bF'' = 0$: then the case $\w = 0$ occurs. Henceforth, the governing equation appears to be simpler than in the other cases: $v'''' = 0$, whose solution reads:
\begin{equation}
v(x) = A_0 + A_1 \, x + A_2 \, x^2 + A_3 \, x^3 .
\end{equation}
As an example, let us consider boundary conditions \eqref{eq:BC_el} with $\hat{\Gamma}_{\sss{R}} =  \hat{\Gamma}_{\sss{L}} = \hat{\Gamma}$, which yields the following values for the constants:
\begin{equation}
A_0 = -\frac{\hat{\Gamma}\,L^2}{16 \, \bA} \quad A_1 = 0 \quad A_2 = \frac{\hat{\Gamma}}{4\,\bA}\quad A3 =0 .
\end{equation}
Of course no bifurcated perturbations would occur in the absence  of hyperstress at the boundary.

\subsection{Numerical Examples}
Let us consider a planar lipid membrane at the fixed temperature $T\sim 30^{\circ}$ (see \Fig{fig:fig_dz_energy}). Under this assumption, the use of the experimental data allows for determining the energetic coefficients $\eqref{eq:ai}$ of the local part of the strain energy density as suggested in \cite{Deseri:2013}. These value are reported in \Tab{tab:tab_value_1}, where the values $J^*$, $J_i$ and $J_*$ represent the configuration balanced by the Maxwell stress (see \cite{Coleman:1988}).
\begin{table}[htbp]
\centering
\begin{tabular}{|c|c|}
\hline 
$T \,[^{\circ}C]$ & $\Sigma_{M} [J/m^2] \, \times 10^{-3}$  \\ 
\hline 
30 & 5.923  \\
\hline 
\end{tabular} 
\begin{tabular}{|c|c|c|c|c|}
\hline 
$J_*$ &$J_i$ &$J^*$ & $J_{\max}$ & $J_{\min}$ \\ 
\hline 
1.02539& 1.16667& 1.30794& 1.0851 & 1.24823  \\
\hline 
\end{tabular} 
\caption{Characteristic values of the membrane stretching energy at $T\sim 30^{\circ}$.}
\label{tab:tab_value_1}
\end{table}

\begin{figure}[htbp]
\centering
\includegraphics[width=0.95\columnwidth]{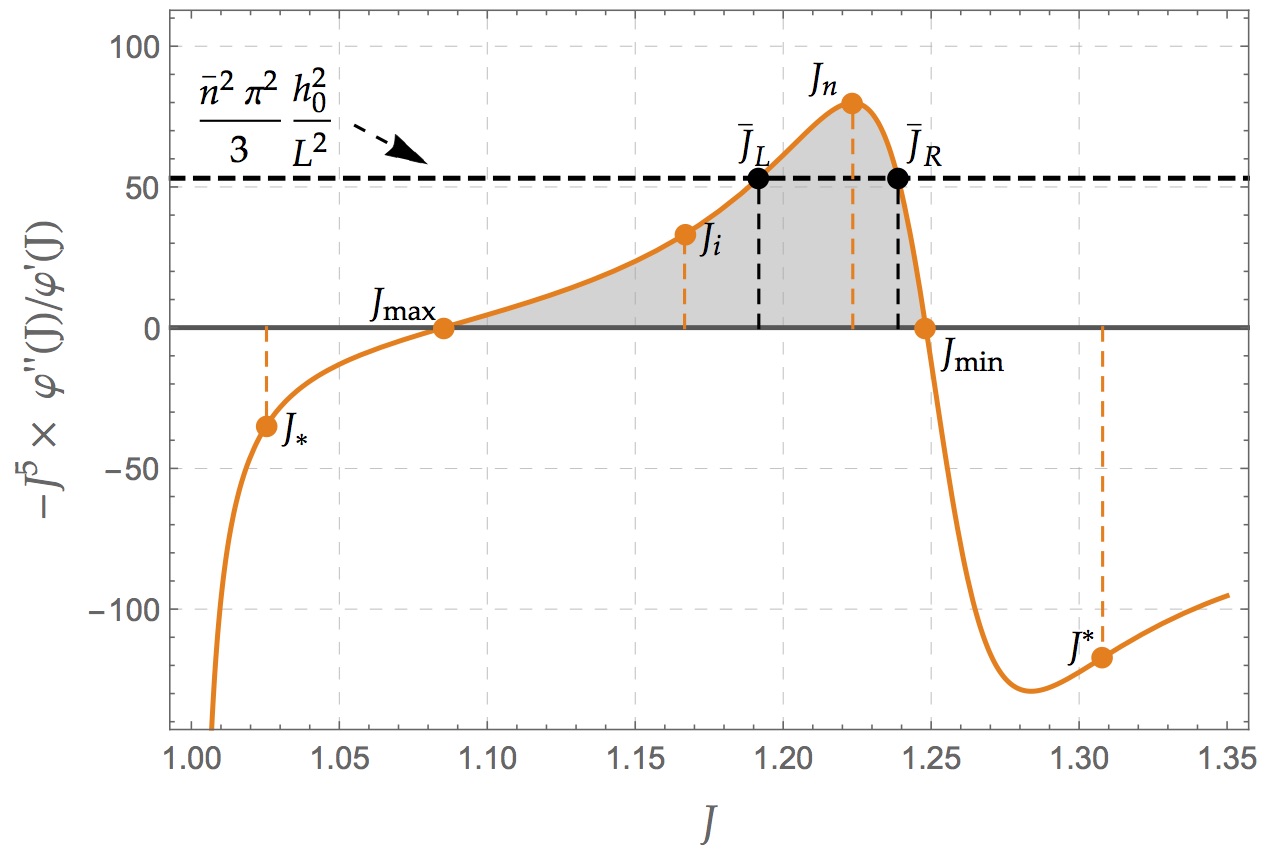} 
\caption{Plot of the ratio  $\bF''/\bF'$ as function of the homogeneous  configuration $\bar{J}$ whenever $n = 8830$, $h_0 = 4.55 \,nm$ and $L = 10 \, \mu m$ (see \eqref{eq:relationship1}).}
\label{fig:fig_elastic_ratio}
\end{figure}

The solution of the problem in \eqref{eq:euler_elastic_2} depends on the sign of the ratio $\bF''/\bF'$, appearing in \eqref{eq:e_f2_f1_1} and \eqref{eq:e_f2_f1_2}. We recall that bifurcations occur if $\bF'' < 0$, i.e. whenever the membrane stretch $\bar{J}$ lies in the unstable part of the spinoidal zone. This circumstance is highlighted in \Fig{fig:fig_elastic_ratio} as a grey region under the orange curve which, as expected, is contained in the spinoidal zone between the two turning points for the convexity of $\f$, i.e. in the range $[J_{max}, J_{min}]$.

\begin{figure}[htbp]
\centering
\includegraphics[width=0.95\columnwidth]{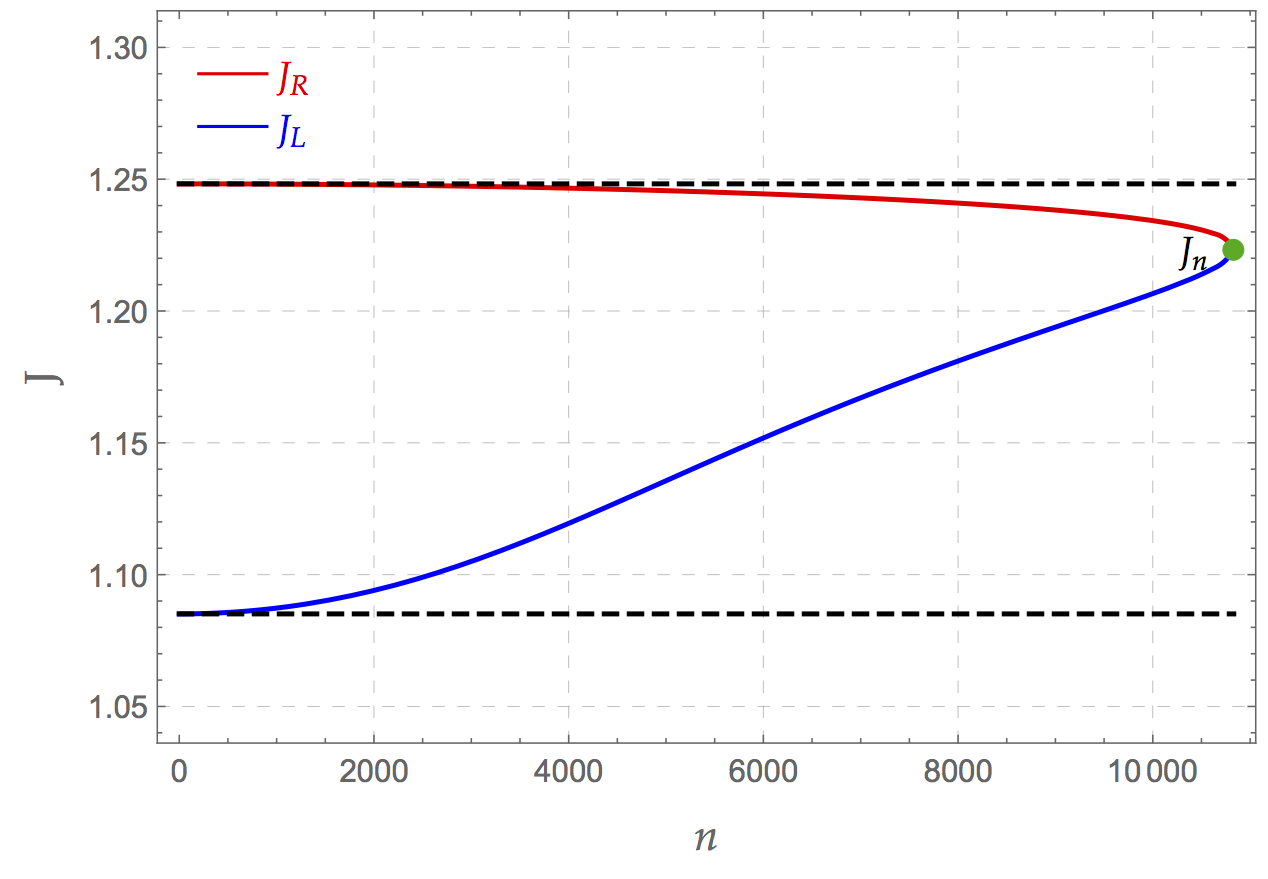} 
\caption{Locus of the \textit{left} and \textit{right} intersections as stretched balanced configurations, i.e. admissible solutions of equation \eqref{eq:relationship1}.}
\label{fig:fig_elastic_JLR}
\end{figure}

\begin{figure}[h]
\centering
\includegraphics[width=0.95\columnwidth]{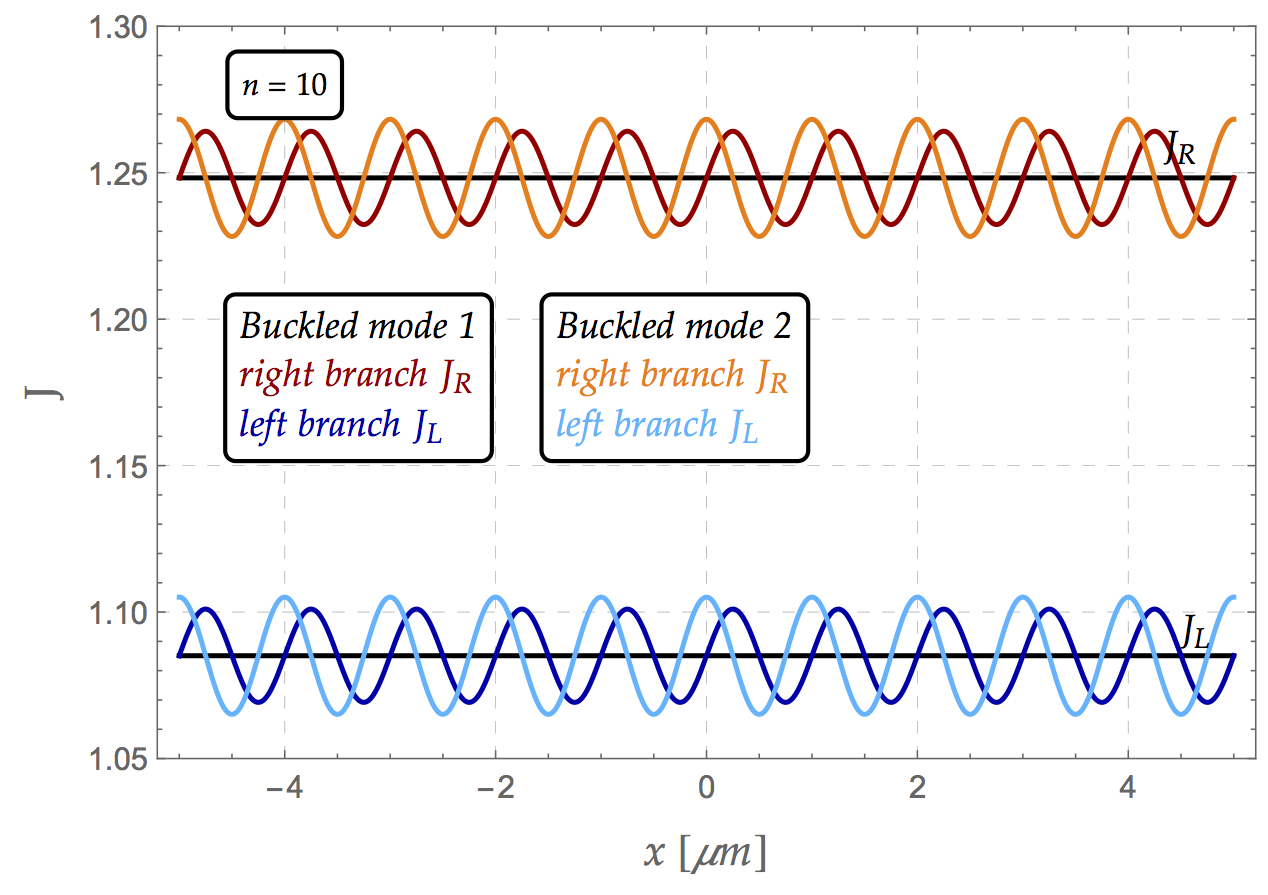} \\
\includegraphics[width=0.95\columnwidth]{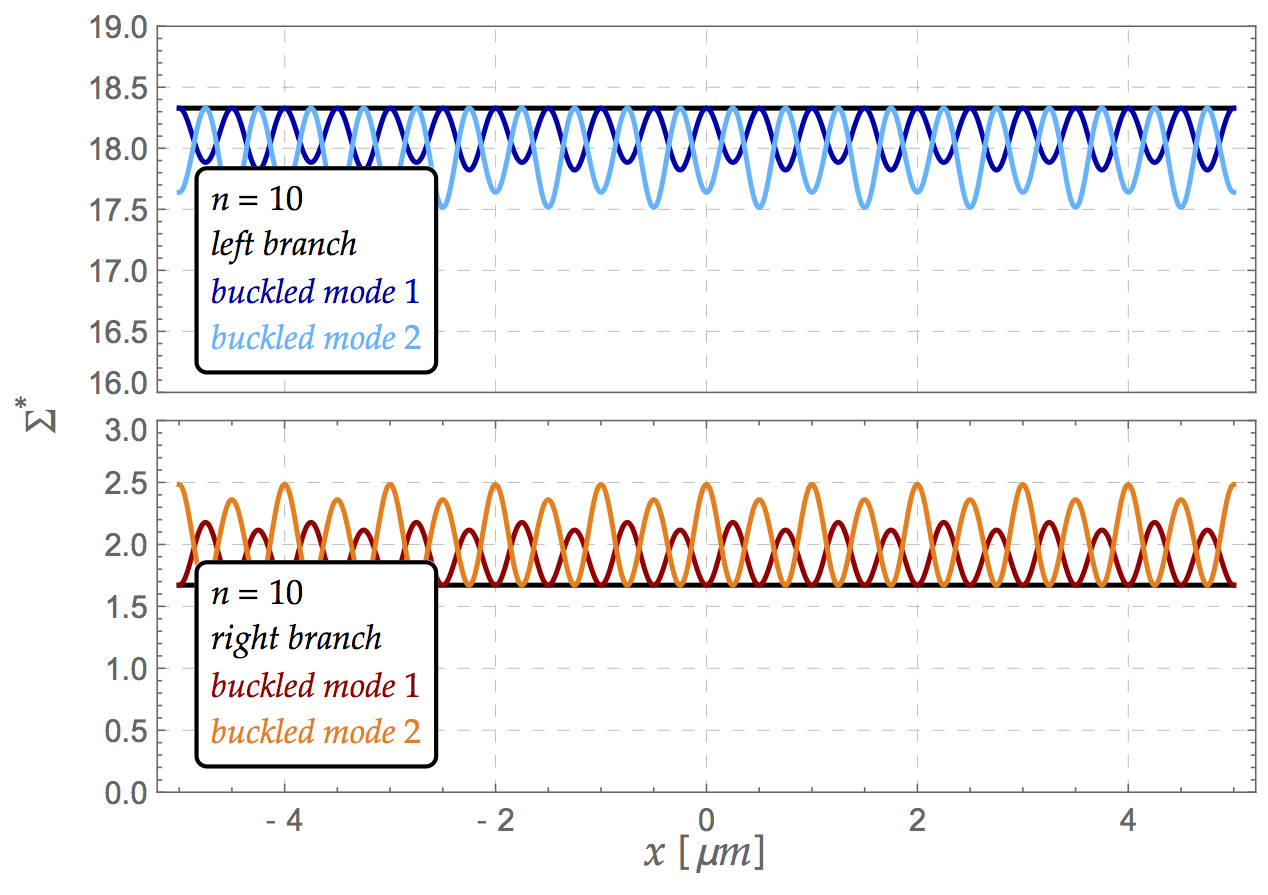}
\caption{\textit{Buckled mode 1}: $A_2 = 0$ and  $\hat{\Gamma} = \frac{\alpha(\bar{J})}{50 \, L}$. \textit{Buckled mode 2}: $A_2 = \frac{1}{50} \frac{L}{2\,\pi\,n}$ and  $\hat{\Gamma} = 0$.}
\label{fig:fig_elastic_modes}
\end{figure}

\Fig{fig:fig_elastic_ratio} shows that for each chosen value of $n$, representing the index mode or ``wave number'', there exist two admissible solutions for \eqref{eq:relationship1}. One of such values of $\bar{J}$ lies on the left and the other one on the right branch of the curve with respect to $J_n$ the location where the horizontal tangent is found. Moreover, this is the only location where a unique value of $n$ is possible, i.e $J_R = J_L = J_n$. The wave number related to this location is labelled $n_{max}$, because \eqref{eq:relationship1} ensures that greater values of $n$ do not allow the presence of bifurcated solutions. This value can be computed as follows:
\begin{equation}
n_{max} =  \frac{1}{\pi}\left(\frac{L}{h_0} \right) \sqrt{-3\frac{\bF'' \bar{J}^5}{\bF'}} \,\Bigg|_{\bar{J}=J_n}.
\end{equation}
The energy used for this numerical example leads to $J_n = 1.2235$ and $n_{max} =  10.832$. Each choice of $n$, therefore, allows for finding two configurations $\bar{J}$ from which a bifurcated mode can be nucleated. Such values are found numerically by choosing values of $n$ from 0 up to $n_{max}$ and computing the intersection $\bar{J}_{\scriptscriptstyle L}$ and $\bar{J}_{\scriptscriptstyle R}$ by means of equation \eqref{eq:relationship1}; The results are shown in \Fig{fig:fig_elastic_JLR}.
The lower blue curve represents the intersection with the left branch of the curve in \Fig{fig:fig_elastic_ratio}, whereas the red curve is the intersection with the right branch. Obviously, these two curves share a common point at $J = J_n$. In order to show the behavior of the system, a value $n = 10$ is chosen for the sake of representation, then the stretch $J$ and the stress $\Sigma$ related to this specific bifurcated configuration are computed through \eqref{eq:equilibrium_dz}. Two cases are considered for illustrative purpose only, in order to show the behavior of our numerical solution: as first case, the arbitrary constant $A_2$ is set to 0 and the hyperstress is chosen such that $\hat{\Gamma} = \frac{\alpha(\bar{J})}{50 \, L}$, whereas in the second example a case with a null hyperstress, $\hat{\Gamma} = 0$, is considered and the constant $A_2$ is chosen as $A_2 = \frac{1}{50} \frac{L}{2\,\pi\,n}$. Both the results are shown in \Fig{fig:fig_elastic_modes}.

\section{The mechanics of fractional order lipid bilayers}
\label{chap:viscous}
Available experimental data \cite{Harland:2010, Espinosa:2011, Craiem:2010} show that lipid bilayers present a time-dependent behavior depending on the in-plane anomalous viscous behavior exhibited by various lipid molecules at different temperatures.

In this section we aim to introduce the governing equations of a Fractional Hereditariness capturing the evolution of the perturbations on the ordered/disordered phase transition shown by lipid bilayers. Such perturbation are predicted to occur in the lipid membrane starting from a homogenously squeezed configuration.

\black{The experimental data about lipid membrane hereditariness that can be found in literature \cite{Harland:2010} show that the case of  a purely elastic membrane represents the asymptotic condition of the mechanics of the lipid bilayer under a constant uniform stress. However, this circumstance is very seldom present in the physiological conditions of living cells, for which intracellular and/or extracellular fluids contributes to change the areal membrane stretch several times during cell lifetimes. Therefore the membrane stress at a certain observation time $t$ may be much higher than the value evaluated in the non-linear elasticity framework, it may evolve into breakage of the cell membrane or to lipid phase modification towards ceramid phase and then to cell apoptosys \cite{Craiem:2010}.
}

The case of the non-homogeneous reference configuration is addressed as in \textbf{Section 2} and it will be not studied in this context for brevity.

\subsection{The physical description of lipid membrane hereditariness }
The mechanics of lipid bilayers forming artificial and natural cytoplasmatic membranes presents a significative hereditary behavior \cite{Espinosa:2011}.  Storage and loss moduli $G^{'}(p)$, $G^{''}(p)$ of lipid membrane depend on the type of lipids (in the membrane e.g. phosphatidylcholine (PODC), the sphyingomyelin (SM) based lipid chains) and on the melting temperatures of such mixtures \cite{Espinosa:2011}. The morphology of the lipids in the bilayers influence their viscosity. It may  be either liquid-ordered or gel-phase, for temperatures over or below the melting temperatures of the PODC. For SM the liquid-disordered or the solid phase (ceramide) may be involved depending upon the temperature of the membrane. 

Several experimental observations on lipid mono-and-bilayer \cite{Harland:2010} showed that the storage and loss modulus, namely $G^{'}(p)$ and $G^{''}(p)$, are proportional to the frequency through a power-law of frational order, i.e $G^{'}(p)\propto p^{\beta}$ and $G^{''}(p)\propto p^{\beta+1}$, where the exponent  $\beta$ depends on temperature and specific chemical composition of the biological structure. 

Henceforth, the use of Maxwell rheological elements to model storage and loss moduli of the material does not provide an suitable representation for the behavior of lipid membrane. This is because Maxwell models yield  $G^{'}(p)\propto p$ and $G^{''}(p)\propto p^2$, which are not observed in experimental rheology of such membranes \cite{Espinosa:2011}.

In this context appropriate models of the hereditary behavior of the lipid membranes must contain fractional-order operators models, in which \textit{creep} and \textit{relaxation}  are described as power-laws of real-order, such that  $J(t)\propto t^{\beta}$ and $G(t)\propto t^{-\beta}$, respectively. The time evolution of small perturbations arising in lipid bilayers from homogeneous configurations describing uniform squeezing is here modeled by making use of the  Boltzmann-Volterra superposition integral. In particular, this allows for measuring the stress evolution at a generic location $x$ depending on an applied strain history $\epsilon(x,t)$ as follows:
\begin{equation}
\label{eq:relaxation}
\sigma(x,t)=\dfrac{C_{\beta}}{\Gamma[1-\beta]}\int_{-\infty}^{t} \left(t-\tau \right)^{-\beta}\, \dot{\epsilon}(x,\tau) \,d\tau;
\end{equation}
the right-hand side of this expression is related to the Caputo fractional-order derivative \cite{Caputo:1969, Podlubny:1998, Magin:2010, Samko:1987,Kilbas:2006}, i.e.:
\begin{equation}
\label{fop}
\DD{\beta} f(t) = \dfrac{1}{\Gamma({\beta})}\int_{-\infty}^{t}(t-\tau)^{-\beta}\dot{f}(x,\tau)d\tau .
\end{equation}
A rheological model known as \textit{springpot element} (after Scott-Blair \cite{Scottblair:1974}) is associated to \eqref{fop}. This represents an intermediate behavior among a linear elastic spring and a viscous dashpot that are obtained for $\beta=0$ and $\beta=1$, respectively. 

In the next section the free energy function obtained in \cite{Deseri:2014} for power-law hereditary materials is utilized. Such a free energy will be further specialized to yield the rheological description of the springpot element to handle lipid membrane hereditariness.

%
\subsection{The free energy of hereditary lipid bilayers}
In this section we aim to introduce the governing equations for the evolution of small perturbations of homogeneous configuration of hereditary and planar lipid membranes.

To this aim it is worth bearing in mind that the quadratic form of the free energy in \eqref{eq:quadratic_energy} contains both a local perturbative term, namely $\e(x\,t)$, and a non-local contribution in term of a first order gradient $\e_x(x,\,t)$. As we observe that the free energy function of the purely elastic case is a function of the state variables $\e(x)$ and  $\e_x(x)$, the free energy function in presence of material hereditariness may be assumed as the sum of different contributions related to the local and the non-local state variables (see e.g. \cite{Breuer:1964, DelDes:1996, DelDes:1997, DesGentGold:1999, DesGoldFab:2006, DesGold:2007}).

By looking at purely (nonlinear) elastic contributions, in the previous section the phase transition phenomena describing areal changes of lipid membranes were obtained \cite{Deseri:2008, Deseri:2013}. Time evolution of small perturbation of such configurations are inferred to be modulated by the local and nonlocal stresses $\sigma_{\sss{L}}(x,t)$ and $\sigma_{\sss{N}}(x,t)$ respectively, i.e.
\begin{subequations}
\label{eq:v_sigma_LN}
\begin{align}
&\sigma_{\sss{L}}(x,t) = \int_0^t G_{\sss{L}}(t-\tau) \dot{\e}(x,\tau) \,\dd\tau ,\\
&\sigma_{\sss{N}}(x,t) = \int_0^t G_{\sss{N}}(t-\tau) \dot{\e}_x(x,\tau) \, \dd\tau ,
\end{align}
\end{subequations}
where $G_{\sss{L}}$ and $G_{\sss{N}}$ are the local and nonlocal relaxation moduli (relative to the configuration $\bar{J}$), respectively, defined as follows:
\begin{subequations}
\begin{align*}
&G_{\sss{L}}(t) := \bF'' + f_{\sss{L}}(t) ,\\
&G_{\sss{N}}(t) := 2\bA + f_{\sss{N}}(t) .
\end{align*}
\end{subequations}
Here the following relationship must hold 
\begin{equation}
\lim_{t\to\infty} f_{\sss{L}}(t) = \lim_{t\to\infty} f_{\sss{N}}(t) = 0,
\end{equation}
as the elastic case has to be retrieved as limit. The specific dependence of the functions  $f_{\sss{L}}(t) $ and $ f_{\sss{N}}(t)$ on time depends on the experimental observation of the evolution of the ordered-disordered phase as well as of their transition zone. Motivated by the experimental evidence discussed in the previous Section, in this work a power law relaxation function are used for the description of the decay behavior of both local and nonlocal evolution. In particular, two different decay laws for describing both the local and the nonlocal contribution are assumed. Thus the following relaxation moduli, based on \cite{Deseri:2014}, are considered:
\begin{subequations}
\label{eq:Gmoduli}
\begin{align}
&G_{\sss{L}}(t) := \bF'' + \CL \,t^{-\betaL},\\
&G_{\sss{N}}(t) := 2\bA+ \CN \, t^{-\betaN},
\end{align}
\end{subequations}%
where $\CL$ and $\CN$ represent generalized moduli of the local and nonlocal relaxations, $\betaL$ and $\betaN$ are the decay exponents of the relaxations (for now chosen in the range $[0,1]$). It is worth nothing that the contributions $\bF'' $ and $2\bA$ in \eqref{eq:Gmoduli} come from the third and fourth terms of the linearized functional in \eqref{eq:en_functional}. The use of an additive relaxation form in \eqref{eq:Gmoduli} corresponds to the use of a fractional order rheological element introduced in \eqref{fop}.

After these considerations, the free energy function $\Psi (x,\,t) $ can be thought as composed by two distinguished contributions:
\begin{equation}
\Psi (x,\,t) = \Psidz(x,\,t) + \Psi_{\sss{V}} (x, \,t),
\end{equation}
where $\Psidz(x,\,t)$ is defined by \eqref{eq:energy_dz} and represents the  elastic contribution to the free energy at equilibrium (see \cite{DelDes:1996}), while $\Psi_{\sss{V}}(x,\,t)$ denotes the free energy associated to the hereditary response of the membrane. This has been shown \cite{Deseri:2014} to be the Staverman-Schartzl  energy \cite{Breuer:1964,DelDes:1996,DelDes:1997}. This result and equations \eqref{eq:v_sigma_LN}, \eqref{eq:Gmoduli} suggest that $\Psi(x, \,t)$ may be written also as:
\begin{equation}
\Psi(x,\,t)  = \Psi_{\sss{L}}(\e(x,\,t)) + \Psi_{\sss{N}}(\e_x(x,\,t)),
\end{equation}
where a local and nonlocal term are accounted for. The former depends on the stretch itself, while the latter on its  gradient. Following \cite{Breuer:1964, Deseri:2014} we introduce a kernel $K(\circ,\circ)$ as a symmetric function, i.e  $K(\circ,\circ) \geq 0$ and  $K(\tau_1,\tau_2) =  K(\tau_2,\tau_1)$. The contribution above can finally be written as follows:
\begin{subequations}
\label{eq:energy_PSI_L_NL}
\begin{align}
& \begin{aligned}
&\Psi_{\sss{L}}(x,\,t)  = \frac{1}{2} K_{\sss{L}} (0,0) \e(x,t)^2   \\
&+\e(x,t) \int_{-\infty}^t \dot{K}_{\sss{L}}(0,t-\tau) \e(x,\tau) \dd \tau \, \\
& +\frac{1}{2}\int_{-\infty}^t \int_{-\infty}^t \ddot{K}_{\sss{L}}(t-\tau_1, t- \tau_2)\e(x,\tau_1) \e(x,\tau_2) \dd \tau_1 \dd \tau_2  ,
\end{aligned} 
\\
& \begin{aligned}
&\Psi_{\sss{N}}(x,\,t) = \frac{1}{2} K_{\sss{N}} (0,0) \e_x(x,t)^2  \\
&+\e_x (x,t)\int_{-\infty}^t \dot{K}_{\sss{N}}(0,t-\tau) \e_x(x,\tau) \dd \tau \,+ \\
& +\frac{1}{2}\int_{-\infty}^t \int_{-\infty}^t \ddot{K}_{\sss{N}}(t-\tau_1, t- \tau_2)\e_x(x,\tau_1) \e_x(x,\tau_2) \dd \tau_1 \dd \tau_2  ,
\end{aligned} 
\end{align}
\end{subequations}
where 
\begin{subequations}
\begin{align}
&K_{\sss{L}}(t,0) := \bar{\f}'' + \frac{C_{\sss{L}}}{\Gamma(1-\betaL)} (t+\d)^{-\betaL} = G_{\sss{L}}^{\d}(t) ,\\
&K_{\sss{N}}(t,0) :=2 \bar{\alpha} + \frac{C_{\sss{N}}}{\Gamma(1-\betaN)} (t+\d)^{-\betaN} = G_{\sss{N}}^{\d}(t),
\end{align}
\end{subequations}
where $\d$ is a preloading time. Of course $K_{\sss{L}}(0,t) = K_{\sss{L}}(t,0)$ and $K_{\sss{N}}(0,t) = K_{\sss{N}}(t,0)$. It is worth noting that the form of equation \eqref{eq:energy_PSI_L_NL} comes from the definition of the Staverman-Schartzl energy \cite{Deseri:2014, DelDes:1996, Breuer:1964}. This result, together with \eqref{eq:Gmoduli}  and the considerations addressed in equations \textbf{(17-22)} by Deseri et al. \cite{Deseri:2014}, allows for writing down the \textit{final} form of the free energy as:

\begin{subequations}
\begin{align}
&\begin{aligned}
&\Psi_{\sss{L}}(x,\,t)  = \frac{1}{2} G_{\sss{L}}^{\d}(0) \e^2(x,t) \\
&+ \e(x,t) \int_{-\infty}^{t} \dot{G}_{\sss{L}}^{\d}(t-\tau) \e(x,\tau)\dd \tau  \\
& + \frac{1}{2}  \int_{-\infty}^{t} \int_{-\infty}^{t} \ddot{G}_{\sss{L}}^{\d}(2t-\tau_1-\tau_2) \e(x,\tau_1) \e(x,\tau_2)\dd \tau_1 \dd \tau_2 ,
\end{aligned} \\
&\begin{aligned}
&\Psi_{\sss{N}}(x,\,t)  = \frac{1}{2} G_{\sss{N}}^{\d}(0) \e_x^2(x,t) \\
&+ \e_x(x,t) \int_{-\infty}^{t} \dot{G}_{\sss{N}}^{\d}(t-\tau) \e_x(x,\tau)\dd \tau  \\
& +\frac{1}{2}  \int_{-\infty}^{t} \int_{-\infty}^{t} \ddot{G}_{\sss{N}}^{\d}(2t-\tau_1-\tau_2) \e_x(x,\tau_1) \e_x(x,\tau_2)\dd \tau_1 \dd \tau_2,
\end{aligned} 
\end{align}
\end{subequations}
where $\e (x,\,t) = v_x (x,\,t)$, where $v (x,\,t)$ represents the perturbation of the configuration of the lipid membranes at the location $x$ and time $t$. Finally, the total (Gibbs) free energy related to the perturbation $v(x,t)$ can be computed as:
\begin{equation}
\label{eq:v_energy}
\begin{aligned}
\scrE  &=  B\, \int_{t_1}^{t_2} \left( \intO \left[\Psi_{\sss{L}}(x,\,t)  + \Psi_{\sss{N}}(x,\,t)\right] \dd x \right) \dd t \\
& \quad - B\,\dO{\Sigma \, v(x,t) + \Gamma \, v_x(x,t)} ,
\end{aligned}
\end{equation}
where $t_1$ and $t_2 > t_1$ are two subsequent times during which the time evolution of the membrane is investigated.

\section{Linearized evolution of lipid membranes}
The governing equation for the evolution of lipid membrane is sought for $v$ by stationarity of the functional $\scrE$ in the class of syncronous variations, i.e. $\d v(\circ,t_1) = \d v(\circ,t_2) $. The computation of the first variation of \eqref{eq:v_energy} (see \ref{chap_A1} for details) leads to the Euler-Lagrange equation in the form:
\begin{equation}
\label{eq:v_EL}
2\bA \, \dfrac{\pd^4}{\pd x^4} \left(v +  \bCN \DD{\betaN} v \right) - \bF'' \,\dfrac{\pd^2}{\pd x^2} \left(v +  \bCL \DD{\betaL} v \right) = y(x) ,
\end{equation}
where $\bCL = \CL/\bF''$  and $\bCN = \CN/2\bA$  represent the normalized local and nonlocal moduli of the membrane, respectively, and the \textit{forcing term} $y(x)$ is defined as follows:
\begin{equation}
 y(x) = 2\bA\dfrac{\pd^4 \, v_0}{\pd x^4} -  \bF'' \dfrac{\pd^2 \, v_0}{\pd x^2} .
\end{equation} 
Here $v_0(x)$ represents an initial perturbation displacement that can be induced on the membrane at the beginning of the observation time, and it can be thought as the initial configuration before the relaxation.The governing equation \eqref{eq:v_EL} is coupled with the following boundary conditions:
\begin{subequations}
\begin{align}
&\left\{
\begin{array}{l}
\text{either} \\
\bF'' \,\left(v' +  \bC_{\sss{L}} \DD{\betaL} v' \right)  - 2\bA \, \left(v '''+  \bC_{\sss{N}} \DD{\betaN} v''' \right)  = \Sigma + \Sigma_0 \\ 
\text{or}  \\
\d v = 0 
\end{array} 
\right. \\
%
&\left\{
\begin{array}{l}
\text{either}  \\
2\bA \,  \left(v'' +  \bC_{\sss{N}} \DD{\betaN} v'' \right) = \Gamma + 2\bA \, \e_0' \\ 
\text{or} \\
\d v '= 0 
\end{array} 
\right.
\end{align}
\end{subequations}
It is worth nothing that the term $\Sigma_0 :=\bF'' \e_0  + 2\,\bA \,\e_0''  $ can be interpreted as the \textit{initial stress} acting on the membrane to hold it in the initially perturbed configuration. Of course, if no initial perturbation is induced on the membrane, equation \eqref{eq:v_EL} and its boundary conditions lead to an eigenvalue problem, examined in Section \ref{chap:complete_equation} in the sequel.

The structure of the linear partial differential equation \eqref{eq:v_EL} allows for separation of variables for the perturbation $v(x,t)$, i.e.:
\begin{equation}
\label{eq:separation}
v(x,t) = f(x) \, q(t) ,
\end{equation}
where $q(t)$ describes the time change of the perturbation or ``transfer function'', and  $f(x)$ describes the shape of the mode. Henceforth, the governing equation can be written in the following form:
\begin{equation}
\label{eq:fde_sep}
\frac{2\bA}{\bF''}\frac{f^{iv}(x)}{f''(x)} = \frac{q(t)+ \bCL \,\DD{\betaL} q(t)}{q(t) + \bCN \,\DD{\betaN} q(t)} = k^2,
\end{equation}
where $k^2$ is a constant to be determined. In this context, relationship \eqref{eq:omega} holds. In this work we are interested in exploring conditions from which oscillations can occur, henceforth only the case $\bF''<0$ is studied. Then:
\begin{equation}
\label{eq:v_sep_var}
-\frac{1}{\w^2}\frac{f^{iv}(x)}{f''(x)} = \frac{q(t)+ \bCL \,\DD{\betaL} q(t)}{q(t) + \bCN \,\DD{\betaN} q(t)} = k^2 ,
\end{equation}
as oscillatory perturbations are explored. In analogy with \eqref{eq:BC_el} the following boundary conditions are assumed for all times $t$:
\begin{equation}
\begin{array}{ll}
\left\{
\begin{aligned}
&v \Big|_{\partial \Omega^-} = v\Big|_{\partial \Omega^+} = 0 \\
& 2\bA \left[v'' + \bCN \, \DD{\betaN}v''\right] \Big|_{\partial \Omega^-}=  2\bA \left[v'' + \bCN \, \DD{\betaN}v''\right] \Big|_{\partial \Omega^+} = \hat{\Gamma} \
\end{aligned}
\right.
\end{array} 
\end{equation}
which by \eqref{eq:separation} imply:
\begin{equation}
\label{eq_v_BC_fode}
\left\{
\begin{aligned}
& f(x)\Big|_{\partial \Omega} = 0 \\
& 2\bA f'' \left[q(t) + \bCN \, \DD{\betaN}q(t)\right] \Big|_{\partial \Omega} = \hat{\Gamma} 
\end{aligned}
\right.
\end{equation}

\subsection{Solution of the space-dependent equation} 
The space-dependent function $f(x)$ is found through \eqref{eq:fde_sep} to obey the following ordinary differential equation:
\begin{equation}
\label{eq:v_fde_time}
f^{iv}(x) + k^2\, \w^2 f''(x) = 0 .
\end{equation}
After setting
\begin{equation}
\label{eq:zeta}
\zeta^2 = k^2 \, \w^2 \, ,
\end{equation}
bearing in mind that $\bF'' < 0$, the solution of \eqref{eq:v_fde_time} reads as 
\begin{equation}
f(x) = A_1 \cos\left(\zeta \, x \right) + A_2 \sin\left(\zeta \, x \right) + A_3 x + A_4 .
\end{equation}
As usual, the boundary conditions \eqref{eq_v_BC_fode} must be used in order to determine the coefficients $A_i$, $i = 1 \div 4$. In particular, a closer analysis of the condition on the second derivative of the space-dependent function $f(x)$ yields:
\begin{equation*}
2\bA \,f''\Bigl|_{\pd \Omega}\left[q(t) + \bCN \,\DD{\betaN} q(t)\right]   = \hat{\Gamma} \qquad \forall \, t .
\end{equation*}
The latter boundary condition can be fulfilled if either $\hat{\Gamma}$ is a prescribed of of time or if it is constant. This second case is explored in the sequel. Whenever $\hat{\Gamma}$ is constant, then
\begin{equation}
\label{eq:v_kn}
q(t) + \bCN\,\DD{\betaN} q(t) = \k_n,
\end{equation}
where $\k_n$ is a constant. Consequently, the boundary condition is written as follows:
\begin{equation}
\label{eq:bc_fde_space}
2\bA \,f''\Bigl|_{\pd \Omega} \, \k_n  = \hat{\Gamma} .
\end{equation}
Moreover, this condition at the edge highlights that the second derivative $v_{xx}(x,\, t) \Big|_{\pd \Omega}$ there can be zero if and only if 
\begin{equation}
\label{eq:v_BC_f''=0}
f''\Bigl|_{\pd \Omega} = 0 \, \Longleftrightarrow \, \hat{\Gamma} = 0
\end{equation}
the hyperstress is zero. For such a case, equation \eqref{eq:v_kn} is irrelevant. Because in this section the attention is focused on the case $\bF'' < 0$, after setting $s = \sin(\zeta L/2)$ and $c = \cos(\zeta L/2)$, the boundary conditions can be written explicitly in the form:
\begin{equation*}
\begin{aligned}
&
\left\{
\begin{array}{l}
A_1 \, c - A_2 \,s - A_3 \dfrac{L}{2} + A_4 = 0 \\
2 \bA \zeta^2 \left(-A_1 \, c + A_2 \, s \right)  \k_n = \hat{\Gamma}
\end{array}
\right.
\, \text{at } x = -\frac{L}{2} \\
&
\left\{
\begin{array}{l}
A_1 \, c + A_2 \,s + A_3 \dfrac{L}{2} + A_4 = 0 \\
2 \bA \zeta^2 \left(-A_1 \, c - A_2 \, s \right) \k_n = \hat{\Gamma}
\end{array}
\right.
\, \text{at } x = +\frac{L}{2}
\end{aligned}
\end{equation*}
Such a  system is the analogue of \eqref{eq:sys_el}:
\renewcommand*{\arraystretch}{1.25}
\begin{equation}
\label{eq:sys_vis}
\left[
\begin{array}{cccc}
0 & s & \frac{L}{2} & 0 \\ 
c & 0 & 0 & 1 \\ 
0 & s & 0 & 0 \\ 
-2\bA \, \k_n \zeta^2 c & 0 & 0 & 0
\end{array} 
\right]
 \, 
\left(
\begin{array}{c}
A_1 \\ A_2 \\ A_3 \\ A_4
\end{array} 
\right)
=
\left(
\begin{array}{c}
0 \\ 0 \\ 0 \\ \hat{\Gamma}
\end{array} 
\right)
\end{equation}
whose nontrivial solutions can be found by studying the roots of the determinant, namely after solving:
\begin{equation}
\bA \, c \, s \, L \, \k_n \, \zeta^2 = 0.
\end{equation}
Because of equation  \eqref{eq:bc_fde_space}, the case $\k_n = 0$ implies that the hyperstress at edges is zero, and for now we do not consider this possibility to occur. Then, the quantities $\bA$, $ L$ and $\k_n$ are always nonzero, and we are left to study only two cases. \\
\quad \\
\textbf{Case 1.} Because $\zeta^2 = k^2\,\w^2$ with $k > 0$ (although still unknown at this stage), if $s = 0$ we have:
\begin{equation}
\label{eq:v_zetasol}
k^2 \, \omega^2 = \frac{4 n^2 \pi^2}{L^2},
\end{equation}
and
\begin{equation}
\label{eq:relationship1_visco}
-\frac{\bF''}{\bF'} \bar{J}^5 = \frac{n^2 \pi^2}{3\,k^2} \left(\frac{h_0}{L} \right)^2.
\end{equation}

\noindent\textbf{Case 2.} If $c = 0$ then $\hat{\Gamma} = 0$. As highlighted in \eqref{eq:v_BC_f''=0}, this happens if and only if $f''\left(\pd \Omega \right) = 0$.

\subsection{Solution of the time-dependent equation} 
The time-dependent solution $q(t)$ turns out to depend on the value of the second derivative in space at the edges (see  \eqref{eq:bc_fde_space}).

Whenever in \eqref{eq_v_BC_fode} the boundary condition on the second derivative of the displacement is nonzero , the presence of a hyperstress $\hat{\Gamma}$ at the edges implies that the time-dependent term is constant, assuring that relation \eqref{eq:v_kn} holds.
This equation can be easily solved through the method of the \textit{Laplace Transform method} (see \ref{chap:A3}) to yield:
\begin{equation}
\label{eq:v_ht1}
q(t) = \frac{\k_n}{\bCN} t^{\betaN} \MLt{\betaN}{\betaN +1}{-\frac{1}{\bCN}t^{\betaN}} + q_0 \MLo{\betaN}{-\frac{1}{\bCN}t^{\betaN}} ,
\end{equation}
where $\MLt{\alpha}{\beta}{z}$ is the Mittag-Leffler function of two parameters. At the same time, the assumption of the separation of variables dictates that \eqref{eq:fde_sep} be satisfied. Hence, \eqref{eq:fde_sep}  and \eqref{eq:v_kn} imply that the following relationship has to hold:
\begin{equation}
\label{eq:fde_sep_2}
 q(t)+ \bCL\, \DD{\betaL} q(t) = k^2\,\k_n,
\end{equation}
whose solution is again found  by using the \textit{Laplace Transform method}:
\begin{equation}
\label{eq:v_ht2}
q(t) = \frac{k^2 \,\k_n}{\bCL} t^{\betaL} \MLt{\betaL}{\betaL +1}{-\frac{1}{\bCL}t^{\betaL}} + q_0 \MLo{\betaL}{-\frac{1}{\bCL}t^{\betaL}} .
\end{equation}
Both equations \eqref{eq:v_ht1} and \eqref{eq:v_ht2} give  explicit analytic closed forms for the time-dependent function $q(t)$. Obviously they must be same. The trivial case in which the local and nonlocal terms have both the same relaxation exponent $\betaL = \betaN$ and the same normalized material parameters $\bCL = -\bCN$ shows that 
\begin{equation*}
k^2 = \frac{\bCL}{\bCN} = 1,
\end{equation*}
bearing in mind that the local term $\bCL< 0$ as it is made dimensioless dividing $\CL$ by $\bF''<0$.

\subsection{Complete time-dependent equation: Eigenvalues}
\label{chap:complete_equation}
\begin{figure}[h]
\centering
\includegraphics[width=0.95 \columnwidth]{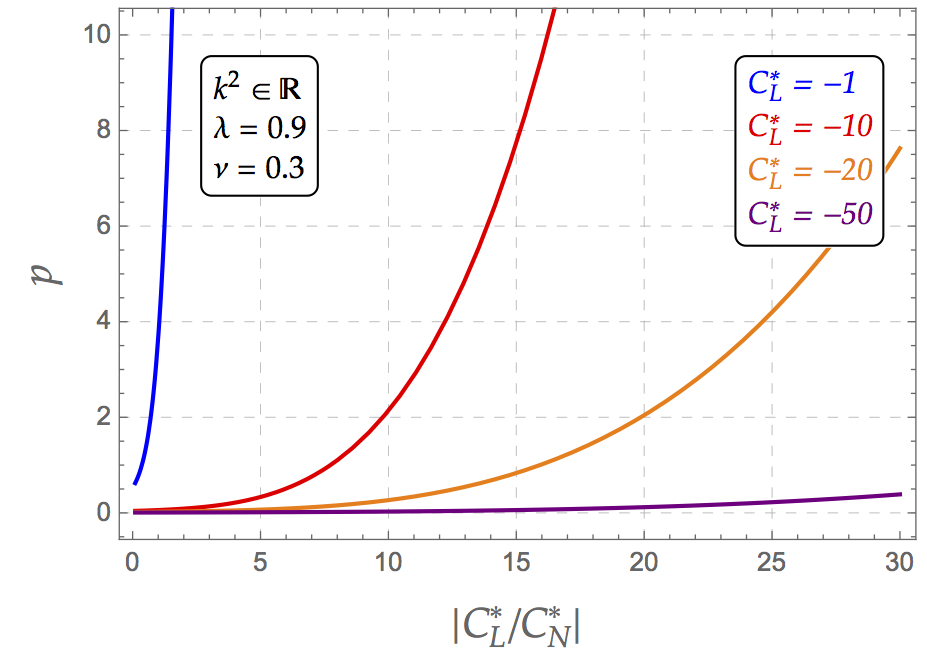} \\
\includegraphics[width=0.95 \columnwidth]{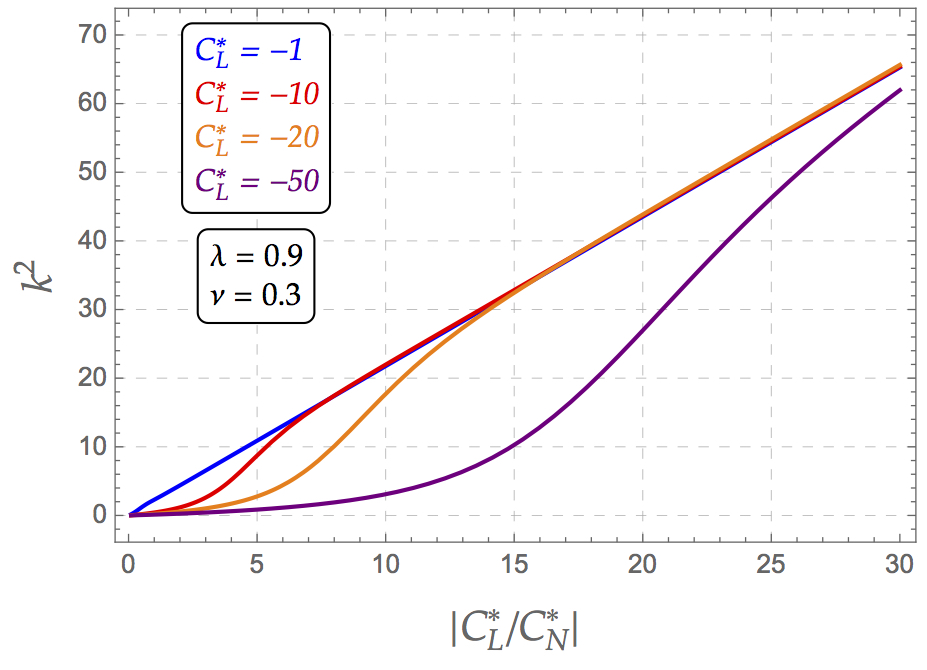}
\caption{Locus of the real eigenvalues $k^2$  and their values as function of the ratio $R = - \bCN /\bCL$ whenever $\betaL = 0.9$ and $\betaN = 0.3$ (see equations \eqref{eq:v_p=0}-\eqref{eq:v_k2}).}
\label{fig:fig_visco_k2_par1}
\end{figure}

\begin{figure}[h]
\centering
\includegraphics[width=0.95 \columnwidth]{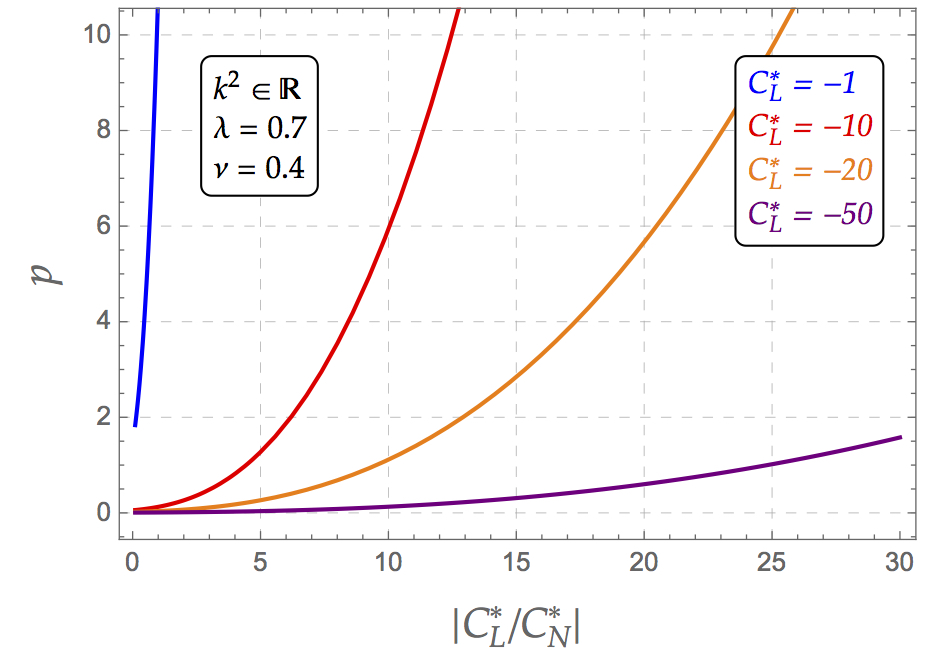} \\
\includegraphics[width=0.95 \columnwidth]{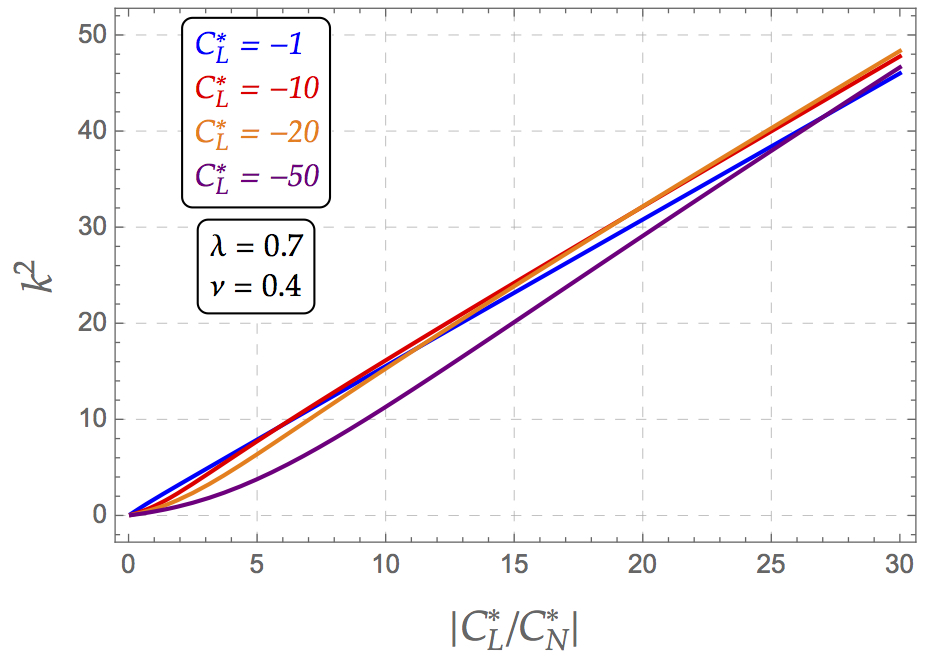}
\caption{Locus of the real eigenvalues $k^2$  and their values as function of the ratio $R = - \bCN /\bCL$ whenever $\betaL = 0.7$ and $\betaN = 0.4$ (see \eqref{eq:v_p=0}-\eqref{eq:v_k2}).}
\label{fig:fig_visco_k2_par2}
\end{figure}

The fact that \eqref{eq:fde_sep} and \eqref{eq:v_kn} must be consistent also in the nontrivial case is studied in this section. In this regard, the complete equation  coming from \eqref{eq:fde_sep} and \eqref{eq:v_kn} is considered:
\begin{equation}
\label{eq:v_fde_time_complete}
\bCL\, \DD{\betaL} q(t)  - \bCN\,k^2\, \DD{\betaN} q(t) +(1-k^2) q(t) = 0.
\end{equation}
Equation \eqref{eq:v_fde_time_complete} has the form of a Fractional Order Eigenvalue Problem, which is not easy to be solved. Indeed, very recent works show the strong effort in finding this kind of solutions \cite{EIG:1,EIG:2,EIG:3,EIG:4,EIG:5}. In order to solve this eigenvalue problem, we make use of the right-sided Fourier transform $Q(p)$
\begin{equation}
Q(p) := \int_0^{+\infty} e^{-i\,p\,t} q(t) \, dt \qquad p \in \mathbb{R} . 
\end{equation}
By Fourier transforming both sides of \eqref{eq:v_fde_time_complete} we obtain: 
\begin{equation}
\label{eq:v_H_fourier}
\left[ \bCL \,(-i\,p)^{\betaL}   - \bCN \, k^2\,(-i\,p)^{\betaN} + (1-k^2)  \right] \, Q(p) = 0.
\end{equation}
The roots of the function inside square brackets supplies the eigenvalues of the fractional differential equation \eqref{eq:v_fde_time_complete}. Consider $-i = e^{-i \frac{\pi}{2}}$ and expand \eqref{eq:v_H_fourier}:
\begin{equation}
\bCL\,p^{\betaL}\, e^{-i \frac{\pi}{2}\, \betaL} - k^2 \bCN\,p^{\betaN}\, e^{-i \frac{\pi}{2}\, \betaN} +(1-k^2) = 0 .
\end{equation}
The constant $k^2$ introduced  in \eqref{eq:fde_sep} must be a real-valued number. Solving equation \eqref{eq:v_H_fourier} in terms of $k^2$ we get:
\begin{equation*}
\begin{aligned}
k^2 & = \frac{1 + \bCL \, p^{\betaL} \left(c_{\betaL} -  i \, s_{\betaL} \right) }{1 + \bCN \, p^{\betaN} \left(c_{\betaN}  - i\, s_{\betaN} \right) } \\
& = \frac{\left( 1 + \bCL \, p^{\betaL} \, c_{\betaL}  \right) - i \,\left( \bCL \, p^{\betaL} \, s_{\betaL} \right)}{\left(1+ \bCN \, p^{\betaN} \, c_{\betaN} \right) - i \,\left( \bCN \, p^{\betaN} \, s_{\betaN} \right)} = \frac{a - i\,b}{c - i \,d} \\
& = \frac{a - i\, b}{c - i\,d} \frac{c + i\,d}{c + i\,d} = \frac{a \,c + b \, d}{c^2 + d^2 } + i \frac{a \,d - b \, c}{c^2 + d^2 },
\end{aligned}
\end{equation*}
where we set
\begin{equation*}
\left\{
\begin{array}{l}
a = 1+ \bCL \, p^{\betaL} \, c_{\betaL} \\ 
b = \bCL \, p^{\betaL} \, s_{\betaL} 
\end{array} 
\right.
\qquad
\left\{
\begin{array}{l}
c =  1+ \bCN \, p^{\betaN} \, c_{\betaN}   \\ 
d = \bCN \, p^{\betaN} \, s_{\betaN} 
\end{array} 
\right. ,
\end{equation*}
and for the sake of convenience the positions $c_{\alpha} = \cos (\alpha \, \pi/2)$ and $s_{\alpha} = \sin (\alpha \, \pi/2)$ are used. Because of the fact that $k$ is real, the following relationships must hold:
\begin{subequations}
\begin{align}
&k^2 =  \frac{a \,c + b \, d}{c^2 + d^2 } \label{eq:v_eig_relationship_1}\\
&a \, d - b \,c  = 0. \label{eq:v_eig_relationship_2}
\end{align}
\end{subequations}
The latter of these conditions allows for characterizing the value $k^2$ as
\begin{equation*}
\bCN \, p^{\betaN} \, s_{\betaN} - \bCL \, p^{\betaL} \, s_{\betaL} + \bCL  \, \bCN \, p^{\betaL + \betaN} \left( s_{\betaN} c_{\betaL} -  c_{\betaN} s_{\betaL}   \right)= 0.
%
\end{equation*}
Bearing in mind the transformation formulae for the difference of two angles, the relationship \eqref{eq:v_eig_relationship_2} becomes:
\begin{equation}
\label{eq:v_p=0}
\begin{aligned}
\bCN \, p^{\betaN} \, &\sin\left( \betaN \, \frac{\pi}{2}\right)  - \bCL \, p^{\betaL} \, \sin\left(\betaL \, \frac{\pi}{2}\right) + \\
& \qquad + \bCL  \, \bCN \, p^{\betaL + \betaN} \sin\left( \left(\betaN -  \betaL\right)\frac{\pi}{2}\right)  = 0
%
\end{aligned}
\end{equation}
Finally, a relationship for $k^2$ is found in the form:
\begin{equation}
\label{eq:v_k2}
k^2 = \frac{\left(1+ \bCL \, p^{\betaL} \, c_{\betaL} \right) \left(1 + \bCN \, p^{\betaN} \, c_{\betaN}\right) + \left( \bCL \, p^{\betaL} \, s_{\betaL} \right) \left( \bCN \, p^{\betaN} \, s_{\betaN} \right) }{\left(1 + \bCN \, p^{\betaN} \, c_{\betaN}  \right)^2 + \left( \bCN \, p^{\betaN} \, s_{\betaN} \right)^2}.
\end{equation}
Whenever the trivial case $\betaL = \betaN$ and $\bCL = \bCN$ is considered, equation \eqref{eq:v_p=0} has solution $p = 0$, that implies $k^2 = 1$, as noticed qualitatively above. The solution of \eqref{eq:v_k2} is cannot be found in closed form. In \Fig{fig:fig_visco_k2_par1} and  \Fig{fig:fig_visco_k2_par2} some numerical results are represented  whenever  the local modulus $\bCL$, the  nonlocal modulus $\bCN$ and  both the viscoelastic exponents are known. The value of $R$ is defined as function of the moduli ratio $\bCL/\bCN < 0$, showing that the eigenvalues are continuous functions, then for each choice of $R$ is possible to find the correspondent value of $k^2$.

\begin{figure}[h]
\centering
\includegraphics[width=0.95 \columnwidth]{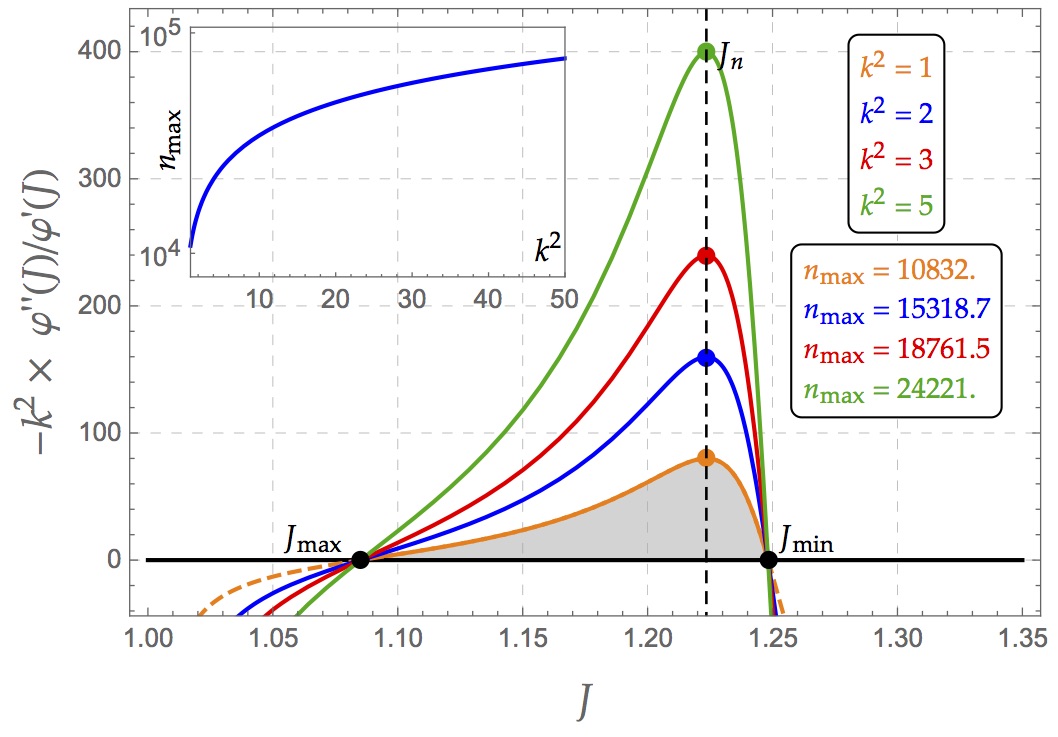}%
\caption{Right hand side of equation \eqref{eq:relationship1_visco} as function of $k^2$.}
\label{fig:fig_visco_k2_ratio}
\end{figure}

\begin{figure}[h]
\centering
\includegraphics[width=0.95 \columnwidth]{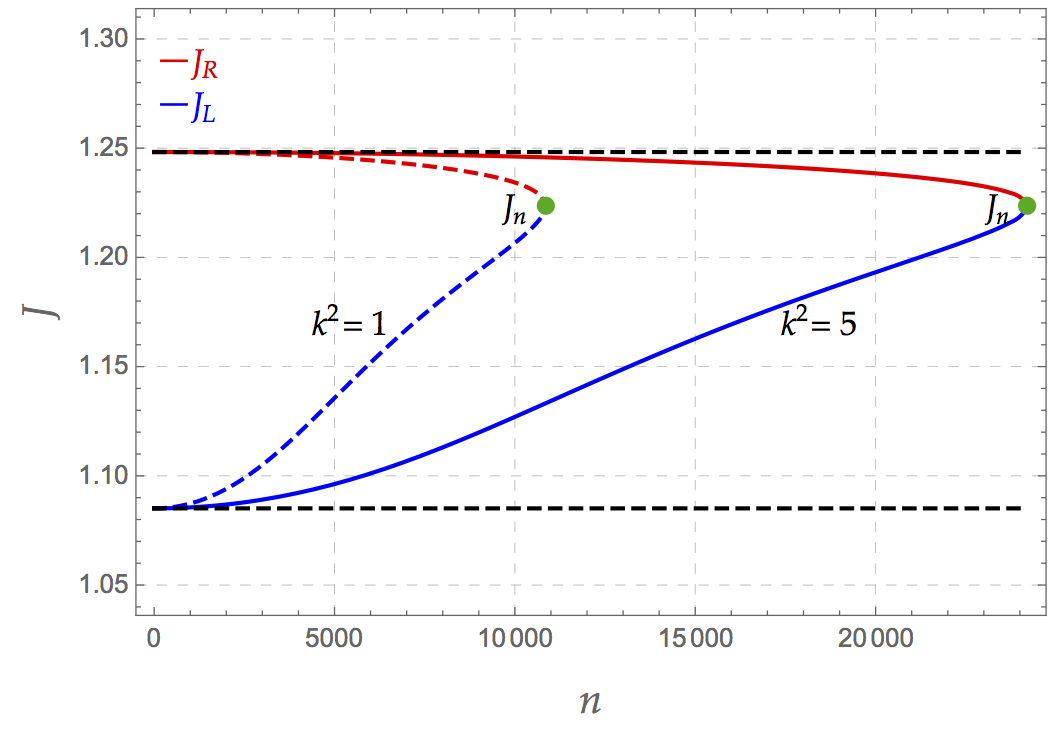}%
\caption{Modification of the \textit{left} and \textit{right} intersection depending on $k^2$ (see also \Fig{fig:fig_elastic_JLR}).}
\label{fig:fig_visco_k2_branches}
\end{figure}

Each bifurcated configuration is characterized by a chosen value of $k^2$ that modifies the left and right branch of the ratio $\bF''/\bF'$, as shown in \eqref{eq:relationship1_visco}. Indeed, the elastic case  \eqref{eq:relationship1} is recovered whenever $k^2 = 1$.
A numerical example handling the same energy used in the elastic case is reported in \Fig{fig:fig_visco_k2_ratio}. A closer analysis of the curves shows that $k^2$ works as a rescaling parameter, increasing the magnitude of the ratio $\bF''/\bF'$ as $k$ increases. The location of $J_n$ is not affected by the rescaling, whereas the upper bound of the curve is deeply influenced by that.  Henceforth, the value $n_{max}$ of the spatial oscillations depends on such a rescaling, as shown in the insert in \Fig{fig:fig_visco_k2_ratio}. Consequently, the left and right branch change their shape, and the intersections yielding the corresponding configurations $\bar{J}$ are modified as shown in \Fig{fig:fig_visco_k2_branches}.

%
\subsection{Initial condition and Eigenvalue problem} 
Let us consider the complete fractional differential equation \eqref{eq:v_fde_time_complete} with inhomogeneous initial conditions:
\begin{equation*}
\left\{
\begin{array}{l}
\bCL\, \DD{\betaL} q(t)  - \bCN\,k\, \DD{\betaN} q(t) +(1-k^2) q(t) = 0 , \\
q(0) = q_0 .
\end{array} 
 \right.
\end{equation*}
As suggested in \cite{Podlubny:1998}, a \textit{Transform method} is used for solving this Fractional Differential Equation. As first step, let use the right-sided Fourier Transform on the original equation taking into account the initial condition:
\begin{equation*}
\begin{aligned}
\bCL \, \Big[ (i\,p)^{\betaL} \hat{Q} &- (i\,p)^{\betaL-1} q_0 \Big] - \bCN\, k^2\,\Big[ (i\,p)^{\betaN} \hat{Q} + \\
&\qquad  - (i\,p)^{\betaN-1} q_0 \Big] + \hat{Q}\,(1-k^2) = 0,
\end{aligned}
\end{equation*}
where $p$ is the variable in the Fourier domain; the solution of the obtained algebraic equation in term of the transformed function $\hat{Q}(p)$ reads:
\begin{equation}
\hat{Q}_k(p) = q_0 \frac{\left(\bCL \, (i\,p)^{\betaL-1} - \bCN \,k^2\, (i\,p)^{\betaN-1} \right)}{\bCL\,(i\,p)^\betaL - \bCN\,k \,(i\,p)^{\betaN} + (1-k^2)} .
\end{equation}
By means of the solution displayed in  \cite{Podlubny:1998}, (eqns 5.22-5.25 pag. 155 where $a = \bCL$, $\beta = \betaL$, $b = -\bCN\,k^2$, $\alpha = \betaN$ and $c = 1- k^2$), the transfer function in the frequency domain of this problem reads as follows:
\begin{equation}
\label{eq:transfer_fun_Gks}
\hat{G}_k(p) = \frac{1}{\bCL\,(i\,p)^\betaL - \bCN\,k^2 \,(i\,p)^{\betaN} + (1-k^2)} .
\end{equation}
It would be worth noting that the transfer function is strictly related to the eigenvalue of $k^2$; for this reason, we denoted $\hat{G}$ with the subscript $k$, in order to highlight the importance of $k^2$ on the transfer function. Finally, the transfer function in the real time domain if found simply by using the Inverse Fourier transform:
\begin{equation}
\begin{aligned}
&G_k(t) = \mathcal{F}^{-1} \left\{ \hat{G}_k(p);t\right\} = \\
&= \frac{1}{\bCL} \sum_{z = 0}^{\infty}(-1)^z \left( \frac{1-k^2}{\bCL}\right)^{z+1} t^{\betaL(z+1)-1} E^{(z)}_{\betaL-\betaN,\betaL+z\betaN}\left(\frac{\bCN}{\bCL}k^2\,t^{\betaL-\betaN} \right) .
\end{aligned}
\end{equation}

The transfer function $G_k(t)$ is strictly connected with the Mittag-Leffler function, and it plays a modulation role in the evolution of the membrane response in terms of both stretch and stress.

\begin{figure}[h]
\centering
\includegraphics[width=0.95 \columnwidth]{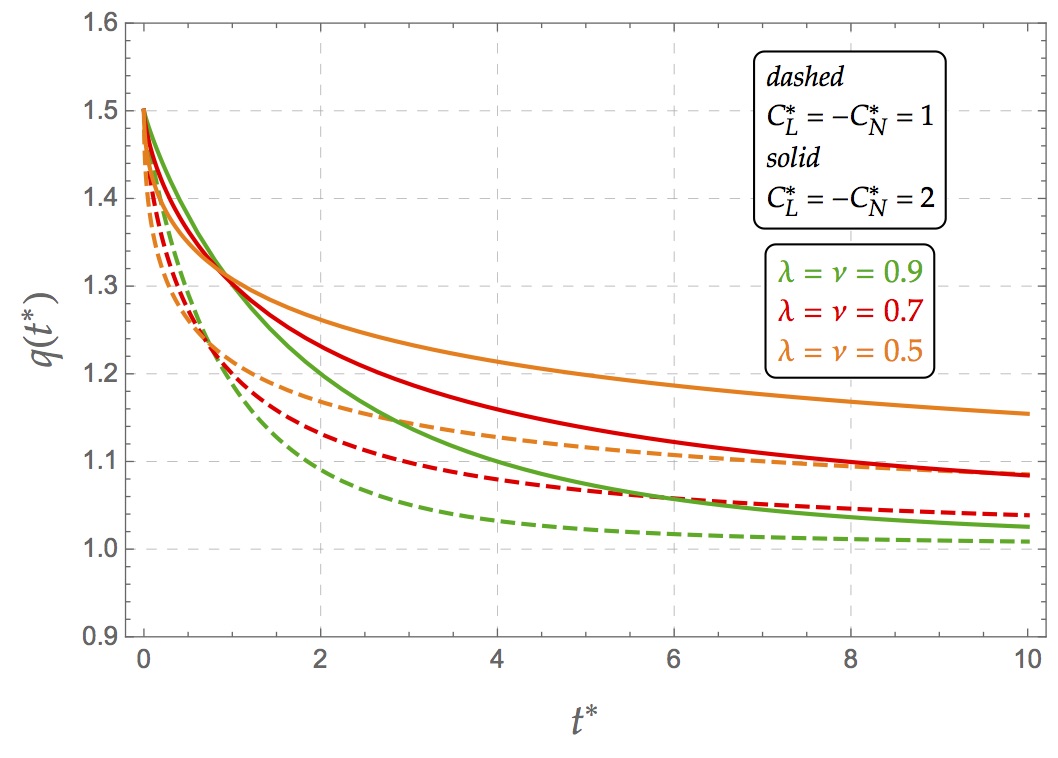}%
\caption{Time-dependent transfer function for two chosen values of  $\bCL = -\bCN $ and $h_0 = 1.5$. Here $t^* = \sqrt[\betaN]{\dfrac{t^{\betaN}}{\bCN}}$ is a dimensionless time (see equation \eqref{eq:v_ht1}.}
\label{fig:fig_visco_example_ht}
\end{figure}

\begin{figure}[h]
\centering
\includegraphics[width=0.95 \columnwidth]{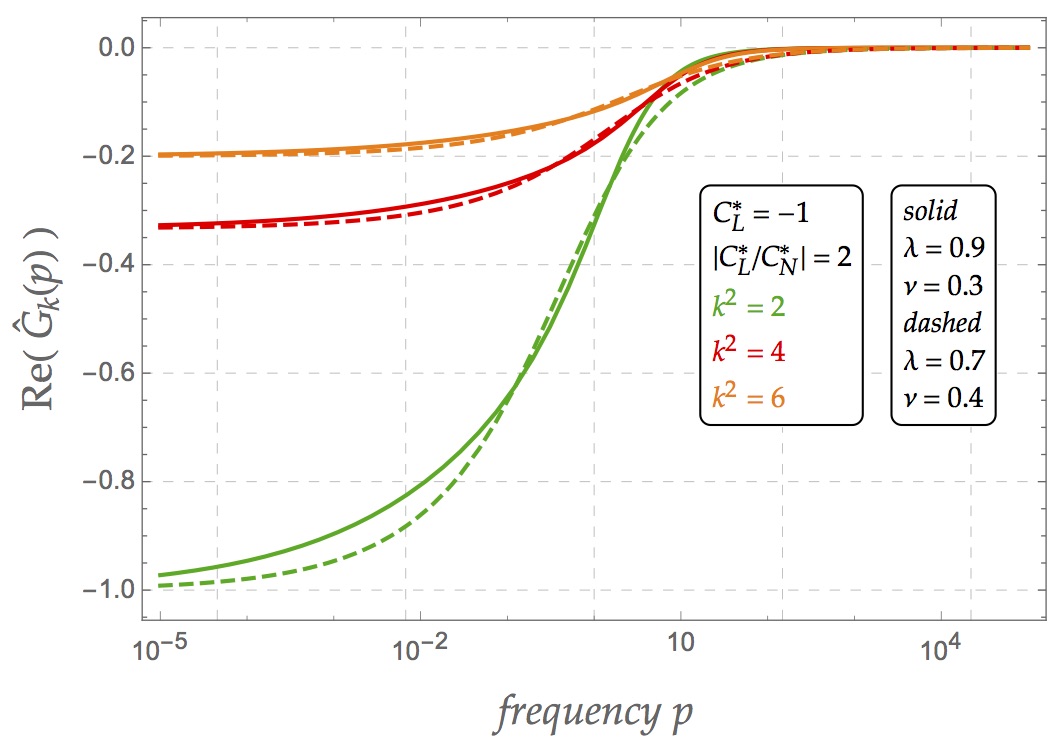} \\
\includegraphics[width=0.95 \columnwidth]{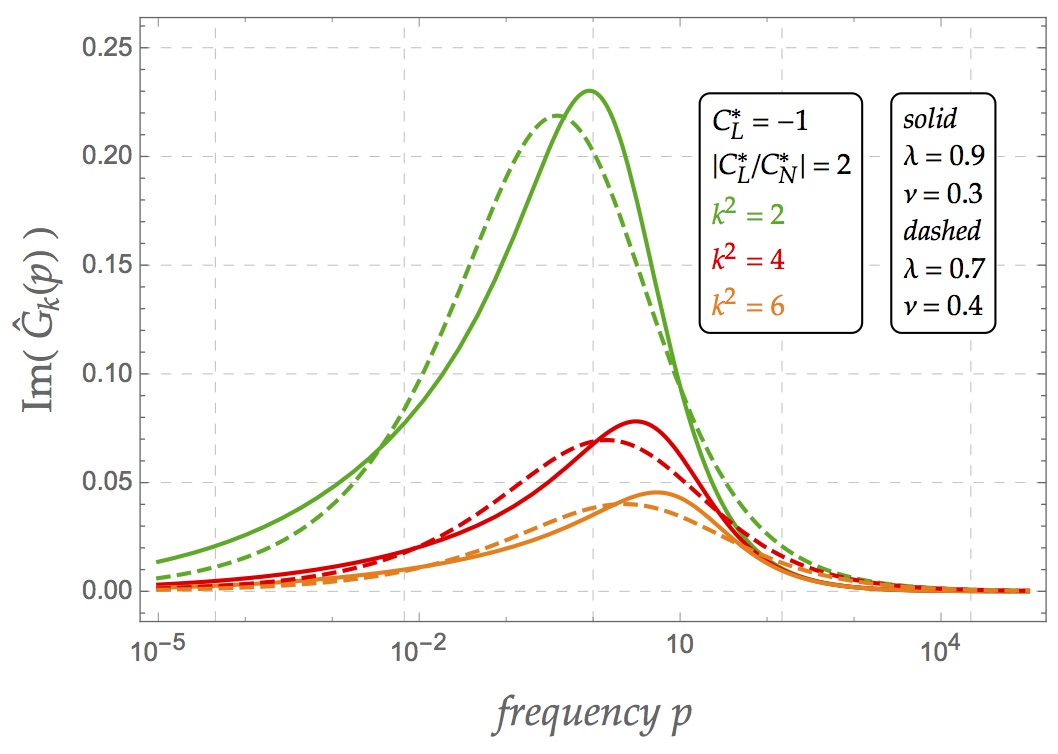}
\caption{Transfer function $\hat{G}_k(p)$: real and imaginary part.}
\label{fig:fig_visco_frequencyplot}
\end{figure}

As an illustrative example, the transfer function is numerically explored in \Fig{fig:fig_visco_example_ht} whenever two subcases of $\bCL = -\bCN $ are considered, by assuming several values of the exponential decay $\betaL = \betaN$. Similarly, in \Fig{fig:fig_visco_frequencyplot} the real and imaginary part of the transfer function are analyzed whenever different exponents of the decay $\betaL \neq \betaN$ are chosen for some values of $k^2$. The Mittag-Leffler function drives the evolution of the membrane stretch, determining changes in the amplitude of the membrane response, as expected from the analysis with a separation of variables.

\section{Discussion and Conclusions}
Lipid phase transition arising in planar membrane and triggered by material instabilities and their linearized evolution are studied in this paper by accounting for the effective viscoelastic behavior inherited by their exhibited power-law in plane viscosity \citep{Espinosa:2011}.

First, the critical set of areal stretches, i.e. the reciprocal of thinning of lipid bilayer, are determined in the limiting case of elasticity and for two sets of boundary conditions. Spatial oscillations corresponding to the nucleated configurations arising from any of such critical stretches are \black{investigated}. Perturbations of the phase ordering of lipids are predicted to form bifurcated shapes, sometimes of large periods relative to the reference thickness of the bilayer. The corresponding membrane stress changes are also oscillatory.

Then, the influence of the effective viscoelasticity of the membrane on its material instabilities is investigated. A variational principle based on the search of stationary points of a Gibbs free energy in the class of synchronous perturbation is employed for such analysis.

The resulting Euler-Lagrange equation is a Fractional Order PDE yielding a non-classical eigenvalue problem. Although its fully general solution is not provided in the paper (see e.g. \cite{EIG:1,EIG:2,EIG:3,EIG:4,EIG:5} for recent analysis on Fractional Eigenvalue Problems), eigenvalues of the viscoelastic problem \eqref{eq:v_fde_time}, namely $\zeta^2$ (see also equation \eqref{eq:zeta}, are found to be amplified by the factor $k^2$ with respect to their elastic counterpart, defined by equation \eqref{eq:euler_elastic_2}. Separation of variables is applied and the mode (spatial dependence) and transfer (time dependence) functions of any admissible perturbations of the stretched configuration are determined.

Time synchronous variations are considered for finding the boundary conditions and the field equations governing the problem. Such equations yield a non-classical eigenvalue problem to be analyzed through the method of separation of variables. Because we analyze bifurcations of the areal stretch from the spinoidal zone, the spatial modes are still found to be oscillatory. The period of oscillation is shown  to decrease with the ratio of (nondimensional) generalized local and nonlocal moduli and, hence, the number of oscillation increase with respect to the elastic case. As the ratio just mentioned above increases, for a given number of oscillations the interval of stretches for which bifurcation can occur gets larger if compared with the one determined by the purely elastic behavior.

First of all, it is found that while the range of critical areal stretches not get affected, the number of oscillations per given critical stretch significantly increase, thereby drastically reducing  the period of oscillations of the bifurcated configurations. Indeed, the factor $k^2$ induces an higher frequency of oscillation. Nevertheless, the relaxation of the bifurcated configurations is shown to occur. For instance, whenever the same power-law applies both for the local and the nonlocal response, the explicit time decay is displayed in \Fig{fig:fig_visco_example_ht}, while in all of the other cases the frequency dependence of the real and imaginary part of the transfer function reveal that fading memory in time occurs as well (see \Fig{fig:fig_visco_frequencyplot}).

The time-dependent part of the problem leads to a non classical fractional eigenvalue problem. Upon exploring the transfer function of the governing equation for different values of the local and nonlocal relaxation power, it can be concluded that time-decay occurs in the response. Hence, large number of spatial oscillation slowly relaxes, thereby keeping the features of a long-tail type response.

Henceforth, although in bifurcated modes a significantly higher number of oscillations is expected than in the limiting case of the equilibrium elastic response of the bilayer, the transfer function, namely the time dependence of bifurcated solution, exhibits a slow decay.
\appendix
\section{Computation of first variation}
\label{chap_A1}
In this Appendix, the explicit calculations referred to functional first variation are displayed.
%
\subsection{Elastic case}
Whenever the elastic case is considered, after neglecting the material constant $B$ the energy functional $\scrE$ in \eqref{eq:en_functional} reads as follows:
\begin{equation*}
\begin{aligned}
\scrE & =  \intO \left(\varphi(\bar{J})+\f'(\bar{J})\,\lambda+\frac{\f''(\bar{J})}{2} \lambda^2+\alpha(\bar{J}) \, \l_x^2 \right) dx  \\
& \quad - \dO{\Sigma \, v +\Gamma \, v_x}
\end{aligned}
\end{equation*}
In this work, the unidimensional case only was considered, then the relationship $ \l= v'(x)$ holds. By entering this result into the energetic functional the first variation reads as follows:
\begin{equation}
\begin{aligned}
\d \scrE  & =   \intO \left( \bF' +\bF'' v' \right) \d v' + (2 \bA \, v'') \d v'' \\ 
& - \dO{\Sigma \d v +  \Gamma \d v'} .
\end{aligned}
\end{equation}
Finally,  after expanding all contributions and integrating by part, the Euler-Lagrange equation with its boundary condition takes the form:
\begin{equation*}
\left\{
\begin{array}{ll}
2 \bA \, v'''' - \bF'' \, v'' = 0   & \text{in} ~\Omega\\ 
\bF'' \, v' - 2\bA \, v''' = \Sigma - \bF ~\text{or}~ \d v = 0 & \text{in}~ \partial \Omega  \\
2 \bA \, v'' = \Gamma  ~\text{or}~ \d v' = 0 & \text{in}~ \partial \Omega  
\end{array} 
\right.
\end{equation*}
%
\subsection{Viscoelastic case}
Whenever the viscoelastic case is studied, we consider the following functional:
\begin{equation}
\begin{aligned}
\EE  = \int_{t_1}^{t_2 } \intO \Big(\psi^{(l)}(\l) & + \psi^{(nl)}(\l_x)  +\\
&  - \dO{\Sigma \, v + \Gamma \, v_x}  \Big) \, \dd x \, \dd t
\end{aligned}
\end{equation}
whose first variation takes the form
\begin{subequations}
\label{eq:v_variation1}
\begin{align}
&\begin{aligned}
\d\EE_{\sss{L}} &= \int_{t_1}^{t_2} \Bigg(\intO \Bigg( G_{\sss{L}}^{\d}(0) \l + \\
& \qquad+\int_{-\infty}^{t} \dot{G}_{\sss{L}}^{\d}(t-\tau) \l(\tau)\dd \tau \Bigg) \d \l \Bigg)\dd x \, \dd t 
\end{aligned}\\
&\begin{aligned} 
\d\EE_{\sss{N}} &= \int_{t_1}^{t_2} \Bigg(\intO \Bigg(G_{\sss{N}}^{\d}(0) \l'  \\
& \qquad + \int_{-\infty}^{t} \dot{G}_{\sss{N}}^{\d}(t-\tau) \l'(\tau)\dd \tau \Bigg) \d \l' \Bigg)\dd x \, \dd t
\end{aligned}
\end{align}
\end{subequations}
Equations \eqref{eq:v_variation1} can be rewritten bearing in mind the Volterra-type integral in the following form:
\begin{subequations}
\begin{align}
& \d\EE_{\sss{L}} = \intO \left(\int_{-\infty}^{t} G_{\sss{L}}^{\d}(t-\tau) \dot{\l}(\tau)\dd \tau\right) \d \l  \,\dd x \\
& \d\EE_{\sss{N}} = \intO \left(\int_{-\infty}^{t} G_{\sss{N}}^{\d}(t-\tau) \dot{\l'}(\tau)\dd\tau\right) \d \l ' \,\dd x 
\end{align}
\end{subequations}
and, after exp substitutions:
\begin{subequations}
\begin{align}
&
\begin{aligned}
\d\EE_{\sss{L}}  = \intO &\Bigg( \bF'' \, \left[ \l(x,t)  - \l_0 \right]  + \\
& \qquad \qquad + C_{\sss{L}} \DD{\betaL} \l (x,t) \Bigg)  \d \l \,\dd x 
\end{aligned}\\
&
\begin{aligned}
\d\EE_{\sss{N}}  = \intO &\Bigg( 2\bA \, \left[ \l'(x,t)  - \l'_0 \right]  + \\
& \qquad \qquad + C_{\sss{N}} \DD{\betaN} \l '(x,t) \Bigg)  \d \l' \,\dd x 
\end{aligned} 
\end{align}
\end{subequations}
Henceforth:
\begin{equation*}
\begin{aligned}
\d \EE & = \intO \Bigg( \left[\bF'' \left(v' - \l_0 \right)  + C_{\sss{L}} \DD{\betaL} v' \right] \d v' + \\
&  \quad \quad +\left[2\bA \left(v'' - \l'_0 \right) + C_{\sss{N}} \DD{\betaN} v'' \right] \d v'' \Bigg) \, \dd x + \\
& \quad - \dO{\Sigma \,\d v +\Gamma \,\delta v'} 
\end{aligned}
\end{equation*}
Finally, the Euler-Lagrange equation reads:
\begin{equation}
\begin{aligned}
&2\bA \, \dfrac{\pd^4}{\pd x^4} \left(v +  C_{\sss{N}}^* \DD{\betaN} v \right) - \bF'' \,\dfrac{\pd^2}{\pd x^2} \left(v +  C_{\sss{L}}^* \DD{\betaL} v \right) =\\
& \qquad \qquad \qquad \qquad= \underbrace{2\bA\dfrac{\pd^2 \, \l_0'}{\pd x^2} -  \bF'' \dfrac{\pd \, \l_0}{\pd x}}_{y(x)}
\end{aligned}
\end{equation}
coupled with the following attendant boundary conditions:
{\footnotesize
\renewcommand*{\arraystretch}{1.5}
\begin{equation}
\left\{
\begin{array}{l}
\text{either} \\
\bF'' \,\left(v' +  C_{\sss{L}}^* \DD{\betaL} v' \right) - \bF'' \l_0  - 2\bA \, \left(v '''+  C_{\sss{N}}^* \DD{\betaN} v''' \right) - 2\bA \l_0''  - \Sigma = 0 \\ 
\text{or}  \\
\d v = 0 
\end{array} 
\right.
\end{equation}
and
\begin{equation}
\left\{
\begin{array}{l}
\text{either}  \\
2\bA \,  \left(v'' +  C_{\sss{N}}^* \DD{\betaN} v'' \right) - 2\bA \l_0' - \Gamma = 0 \\ 
\text{or} \\
\d v '= 0 
\end{array} 
\right.
\end{equation}
}
where the following normalized moduli were used:
\begin{equation}
\bCL = \frac{\CL}{\bF''} \quad  \text{,} \quad \bCN = \frac{\CN}{2 \bA} .
\end{equation}

\section{Computation of the extra energy}
\label{chap_A2}
The $n^{\text{th}}$ buckled mode obtained in the elastic case \eqref{eq:e_modes} can be expanded as follows:
\begin{equation}
\begin{aligned}
&v=\frac{\Gamma\,L^2}{8 \pi^2 \bar{\alpha}}\left[1- \cos \left(\w x\right) \right] + A_2 \sin \left(\w x\right)  \\
&v' =\frac{\Gamma\,L^2}{8 \pi^2 \bar{\alpha}}\, \w\,\sin \left(\w x\right) +   A_2 \, \w\, \cos \left(\w x\right)\\
&v'' = \frac{\Gamma\,L^2}{8 \pi^2 \bar{\alpha}} \, \w^2\,\cos \left(\w x\right)  -A_2 \, \w^2\, \sin \left(\w x\right) 
\end{aligned}
\end{equation}
Because the solution $v(x)$ depends on the trigonometric functions, for the sake of discussion we distinguish two contributions related to the \textit{cosine} and \textit{sine} components, respectively, i.e 
\begin{equation}
v = v_c + v_s .
\end{equation}
The energy stored by the membrane for getting the final configuration from the reference one can be then decomposed as follows:
\begin{equation}
\begin{aligned}
\scrE &= \intO \bF' \,v'(x) + \frac{\bF''}{2} \, v'^2 + \bA \, v''^2 \\
& = \intO \bar{\f}'\left(v_c' + v_s'\right)+ \frac{\bar{\f}''}{2} \left(v_c' + v_s'\right)^2 + \bar{\alpha} \left(v_c'' + v_s''\right)^2 \\
& = \intO \left(\bF' v_c' + \frac{\bF''}{2} v_c'^2 + \bA v_c'' \right) + \left(\bF' v_s' + \frac{\bF''}{2} v_s'^2 + \bA v_s'' \right)  + \\
& \qquad \quad \quad  + 2 \left(\frac{\bar{\f}''}{2} \left( v_c' v_s'\right) + \bar{\alpha} \left(v_c'' v_s''\right) \right) \\
& =\scrE_s + \scrE_c + \scrE_{cs}
\end{aligned}
\end{equation}
where the following relationships were assumed:
\begin{subequations}
\label{eq:e_computed_energy_sc}
\begin{align}
& \scrE_c = \intO \bF' v_c' + \frac{\bF''}{2} v_c'^2 + \bA \,v_c'' , \\
& \scrE_s = \intO \bF' v_s' + \frac{\bF''}{2} v_s'^2 + \bA v\,_s'' , \\
& \scrE_{cs} = 2\intO \frac{\bar{\f}''}{2} \left( v_c' v_s'\right) + \bar{\alpha} \, \left(v_c'' v_s''\right).
\end{align}
\end{subequations}
Let us now compute the energy term by term:
\begin{equation*}
\begin{aligned}
\scrE_c &= \intO \bar{\f}' \,v_c' + \frac{\bar{\f}''}{2} \, v_c'^2 + \bar{\alpha} v_c''^2 \\
& = \w \, \frac{\Gamma\,L^2}{8 \pi^2 \bar{\alpha}}\, \intO \bF' \, \sin \left(n \pi \frac{x}{L}\right) dx+ \\
& \quad + \w^2 \left(\frac{\Gamma\,L^2}{8 \pi^2 \bar{\alpha}}\right)^2\, \intO \left[ \frac{\bF''}{2} \sin \left(n \pi \frac{x}{L}\right)^2  +  \bA\,\w^2\,\cos \left(n \pi \frac{x}{L}\right)^2 \right] dx\\
& =  \w^2 \left(\frac{\Gamma\,L^2}{8 \pi^2 \bar{\alpha}} \right)^2\,\frac{L}{2} \bA \left(\frac{\bF''}{2 \bA} + \w^2 \right) = 0
\end{aligned}
\end{equation*}
since we are studying the case $\w^2 = - \bF''/2\bA$. Analogously
\begin{equation*}
\begin{aligned}
\scrE_s&= \intO \bar{\f}' \,v_s' + \frac{\bar{\f}''}{2} \, v_s'^2 + \bar{\alpha} v_s''^2 \\
& = \w \, A_2\, \intO \bF' \, \cos \left(n \pi \frac{x}{L}\right) dx+ \\
& \quad +\w^2 A_2^2 \intO \left[ \frac{\bF''}{2} \cos \left(n \pi \frac{x}{L}\right)^2  +  \bA\,\w^2\,\sin \left(n \pi \frac{x}{L}\right)^2 \right] dx\\
& =  \w^2 A_2^2\,\frac{L}{2} \bA \left(\frac{\bF''}{2 \bA} + \w^2 \right) = 0,
\end{aligned}
\end{equation*}
and, at least:
\begin{equation*}
\begin{aligned}
\scrE_{cs} = \intO 2 \left(\frac{\bar{\f}''}{2} \left( v_c' v_s'\right) + \bar{\alpha} \left(v_c'' v_s''\right) \right) = 0
\end{aligned}
\end{equation*}
because of the orthogonality of the trigonometric functions. Indeed $v_s'\, v_c' \propto v_s'' \,v_c'' \propto \sin\left(\hat{\w}\,x\right) \cos\left(\hat{\w}\,x\right)$, and since they are orthogonal functions, the integral over a period is zero.

\section{Solution of a Fractional Ordinary Differential Equation}
\label{chap:A3}
As an example, let us solve the following Fractional Order Differential Equation:
\begin{equation}
\label{eq_a_fod}
a \DD{\alpha}h(t) + b\,h(t)= c,
\end{equation}
and denote with $h_0$ the initial condition. The Laplace Transform of \eqref{eq_a_fod} takes the form:
\begin{equation}
\left(a p^{\alpha} + b \right) H= \frac{c}{p} + a\,p^{\alpha-1} h_0
\end{equation}
For the sake of convenience, we distinguish to contribution to the transformed function $H$:
\begin{equation}
\left\{
\begin{aligned}
&H_1 =  \frac{c\,p^{-1}}{a p^{\alpha} + b} = \frac{c}{a} \frac{p^{-1}}{p^{\alpha} + \frac{b}{a}} \\
&H_2 =  \frac{a\,p^{\alpha-1} h_0}{a p^{\alpha} + b} = h_0\frac{a}{a} \frac{p^{-1}}{p^{\alpha} + \frac{b}{a}} 
\end{aligned}
\right.
\end{equation} 
Let us recall the \textit{Laplace Transform of the Mittag-Leffler function} (see Podlubny pag 21, eqn 1.80 \cite{Podlubny:1998})
\begin{equation}
\mathcal{L}\left\{t^{\alpha^* \,k + \beta^* -1} E^{(k^*)}_{\alpha^*,\beta^*} \left(\pm a^* t^{\alpha^*} \right);t;p\right\} = \frac{k^*! \, p^{\alpha^* - \beta^*}}{\left(p^{\alpha^*} \mp a^* \right)^{k^* +1}} ,
\end{equation}
and look for the Anti-transform of each term. For the first term we get recognize that $k^* = 0$, $\alpha^* - \beta^* = -1$, $\alpha^* = \alpha$, $a^* =b/a$, then:
\begin{equation}
h_1(t) = \frac{c}{a} t^{\alpha} \MLt{\alpha}{\alpha +1}{-\frac{b}{a}t^{\alpha}}   ,
\end{equation}
whereas for the second one $k^* = 0$, $\alpha^* - \beta^* =\alpha -1$, $\alpha^* = \alpha$, $a^* =b/a$, then
\begin{equation}
h_2(t) = h_0 t^0 \MLt{\alpha}{1}{-\frac{b}{a}t^{\alpha}}   .
\end{equation}
Finally, the sought solution takes the following form:
\begin{equation}
h(t) = \frac{c}{a} t^{\alpha} \MLt{\alpha}{\alpha +1}{-\frac{b}{a}t^{\alpha}} + h_0 \MLo{\alpha}{-\frac{b}{a}t^{\alpha}} .
\end{equation}

\quad \newline \\
\textbf{Acknowledgements} \\
The authors are grateful to financial support provided in the course of the study. Luca Deseri acknowledges the Department of Mathematical Sciences and the Center for Nonlinear Analysis, Carnegie Mellon University through the NSF Grant No. DMS-0635983 and the financial support from Grant INSTABILITIES - ERC-2013-ADG-``Instabilities and nonlocal multiscale modelling of materials" held by \textit{Prof. Davide Bigoni}, who is gratefully acknowledged. Pietro Pollaci greatly acknowledges the Italian INdAM-GNFM for the financial support through ``Progetto Giovani 2014 – Mathematical models for complex nano- and bio-materials''. Massimiliano Zingales acknowledges the PRIN2010-2011 with national coordinator Prof. A. Luongo. Kaushik Dayal acknowledges support from AFOSR Computational Mathematics (YI FA9550-12-1-0350), NSF Mechanics of Materials (CAREER 1150002), and ONR Applied and Computational Analysis (N00014-14-1-0715).

\quad \newline \\
\textbf{References} \\
\bibliographystyle{elsarticle-num} 
\bibliography{bib_JMBBM.bib} 

\end{document}